\documentclass[aps,twocolumn,pra,superscriptaddress,amsmath,showpacs,tightenlines,pdflatex]{revtex4}

\usepackage{amsmath}

\usepackage{graphicx,times}
\usepackage{amstext}
\usepackage{amsmath}            
\usepackage{amssymb}            
\usepackage{graphicx}           
\usepackage{latexsym}
\usepackage{bm}
 \usepackage{color}

 \def\fig#1{{#1}}

 \def\extra#1{\emph{#1}}
\def \etal{\textit{et al.}}

\newcommand{\Arg}{{\rm arg\,}}
\newcommand{\ket}[1]{|#1\rangle}
\newcommand{\bra}[1]{\langle#1|}
\newcommand{\brkt}[2]{\langle#1|#2\rangle}
\newcommand{\fidelity}[2]{|_d\langle#1|#2\rangle_d|^2}
\newcommand{\fidelityD}[3]{|_{#3}\langle#1|#2\rangle_{#3}|^2}
\newcommand{\NCS}{{$|\alpha\rangle_{d}\,$}}
\newcommand{\LCS}{{$|\beta\rangle_{d}\,$}}

\newcommand{\jpc}{J. Phys.: Condens. Matter~}

\newcommand{\job}{J. Opt. B: Quantum Semiclassical Opt.}

\begin{document}

\title{Phase-space interference of states optically truncated
by quantum scissors:\\ Generation of distinct superpositions of
qudit coherent states by displacement of vacuum}

\author{Adam Miranowicz}
\affiliation{Faculty of Physics, Adam Mickiewicz University,
PL-61-614 Pozna\'n, Poland} \affiliation{CEMS, RIKEN, Saitama
351-0198, Japan}

\author{Ma\l{}gorzata Paprzycka}
\affiliation{Faculty of Physics, Adam Mickiewicz University,
PL-61-614 Pozna\'n, Poland}

\author{Anirban Pathak}
\affiliation{Department of Physics and Materials Science and
Engineering, JIIT, A-10, Sector-62, Noida, UP-201307, India}
\affiliation{RCPTM, Joint Laboratory  of Optics of Palack\'y
University and \\  Institute of Physics of Academy  of Science of
the Czech  Republic, \\ Faculty of Science, Palack\'y University,
771 46 Olomouc, Czech Republic}

\author {Franco Nori}
\affiliation{CEMS, RIKEN, Saitama, 351-0198, Japan}
\affiliation{Physics Department, The University of Michigan, Ann
Arbor, Michigan 48109-1040, USA }

\date{\today}

\begin{abstract}
Conventional Glauber coherent states (CS) can be defined in
several equivalent ways, e.g., by displacing the vacuum or,
explicitly, by their infinite Poissonian expansion in Fock states.
It is well known that these definitions become inequivalent if
applied to finite $d$-level systems (qudits). We present a
comparative Wigner-function description of the qudit CS defined
(i) by the action of the truncated displacement operator on the
vacuum and (ii) by the Poissonian expansion in Fock states of the
Glauber CS truncated at $(d-1)$-photon Fock state. These states
can be generated from a classical light by its optical truncation
using nonlinear and linear quantum scissors devices, respectively.
We show a surprising effect that a macroscopically distinguishable
superposition of two qudit CS (according to both definitions) can
be generated with high fidelity by displacing the vacuum in the
qudit Hilbert space. If the qudit dimension $d$ is even (odd) then
the superposition state contains Fock states with only odd (even)
photon numbers, which can be referred to as the odd (even) qudit
CS or the female (male) Schr\"odinger cat state. This phenomenon
can be interpreted as an interference of a single CS with its
reflection from the highest-energy Fock state of the Hilbert
space, as clearly seen via phase-space interference of the Wigner
function. We also analyze nonclassical properties of the qudit CS
including their photon-number statistics and nonclassical volume
of the Wigner function, which is a quantitative parameter of
nonclassicality (quantumness) of states. Finally, we study optical
tomograms, which can be directly measured in the homodyne
detection of the analyzed qudit cat states and enable the complete
reconstructions of their Wigner functions.
\end{abstract}

\pacs{
 42.50.Gy   
 42.50.Dv,  
 }

\maketitle \pagenumbering{arabic}


\section{Introduction}

Coherent states (CS), since their original introduction by
Schr\"odinger~\cite{Schrodinger26}, Glauber~\cite{Glauber63}, and
Sudarshan~\cite{Sudarshan63}, have been playing a central role in
quantum physics~\cite{Zhang90,Combescure12}, including quantum and
atom optics, mathematical physics, solid-state physics (e.g.,
theories of superconductivity), quantum field theory and string
theory. The conventional infinite-dimensional bosonic CS are the
most classical pure quantum states of the quantum harmonic
oscillator. The importance of CS can be clearly seen through the
Wigner or, equivalently, Glauber-Sudarshan formalisms of quantum
mechanics based on quasiprobabilities in phase
space~\cite{SchleichBook}. Much effort, including the works of
Perelomov~\cite{Perelomov72} and Gilmore (as reviewed in
Ref.~\cite{Zhang90}), has been focused on generalizations of CS
for finite-dimensional bosonic or fermionic CS.

Here we study qudit coherent states (QCS), i.e.,
finite-dimensional analogs of the conventional
infinite-dimensional Glauber
CS~\cite{Buzek92,Kuang93,Miran94,Opatrny95,Leonski97,Marchiolli03,Borzov06,
Mirzaee07,Rezaei09} (for a review see~\cite{Miran01}). QCS were
studied since the 1990s in the context of quantum phase problem
(especially for the Pegg-Barnett formalism, for reviews
see~\cite{Pegg97,Tanas96}), and quantum information and
engineering (reviewed in, e.g.,
Refs.~\cite{Leonski01,DellAnno06,Leonski11}) in qudit systems.

In general, qudit states defined in Hilbert space ${\cal H}^{(d)}$
of a finite dimension $d$ can be expanded in the Fock-state
$|n\rangle$ basis as
\begin{equation}
|\psi\rangle_d=\sum_{n=0}^{d-1}c_{n}|n\rangle,\label{eq:state1}
\end{equation}
where $c_{n}$ are the properly normalized superposition
coefficients. In the special cases for $d=2,3,4$, the states
$|\psi\rangle_d$ are often referred to as the qubit, qutrit, and
quartit (or ququart) states, respectively.

Quantum information processing with qudits has some practical
advantages over that with qubits as it could speed up certain
computing tasks, by simplifying logic
gates~\cite{Muthukrishnan00,Ralph07,Lanyon09}, improving quantum
cryptography~\cite{Bregman08}, and using physical resources more
efficiently~\cite{Yusa05}. Experimental demonstrations of quantum
information processing with qudits include quadrupolar nuclear
spins (i.e. quartits) controlled with nuclear magnetic resonance
in bulk liquids~\cite{Bonk04}, bulk solids~\cite{Kampermann05},
and semiconductor quantum wells~\cite{Yusa05}, as well as
superconducting phase qudits with a number of levels $d$ up to
five~\cite{Neeley09}, and photonic qutrits in linear optical
architectures~\cite{Lanyon09}. Optical qudit states can be
physically generated from infinite-dimensional states by optical
state truncation using the so-called quantum scissors
devices~\cite{Leonski01,Leonski11}. Simple examples of such
scissors are shown in Fig.~1.

Various nonclassical properties of qudit states were investigated
in several occasions. Interestingly, the states are referred by
different names depending on the functional form of $c_{n}.$ We
are in general interested in the nonclassical properties of the
quantum state described by Eq.~(\ref{eq:state1}). However, the
present study would be focused on the QCS.

The conventional infinite-dimensional Glauber CS $\ket{\alpha}$ of
light can be defined in several equivalent ways. For example, (i)
by the action of the displacement operator
$\hat{D}(\alpha,\alpha^*)\equiv \exp(\alpha\hat{a}^{\dagger}
-\alpha^* \hat{a})$ on the vacuum state $|0\rangle$, where
$\hat{a}$ ($\hat{a}^{\dagger}$) is the conventional
infinite-dimensional annihilation (creation) operator.
Equivalently, these CS can be defined by: (ii) $\ket{\alpha}={\cal
N}\exp(\alpha\hat{a}^{\dagger})|0\rangle$ as implied by the
definition (i) but with the displacement operator factorized
according to the Campbell-Baker-Hausdorff theorem. Hereafter, the
function ${\cal N}$ normalizes a given state $\ket{\psi}$, i.e.,
${\cal N} \ket{\psi}=\ket{\psi}/\sqrt{\brkt{\psi}{\psi}}$. These
equivalent definitions of the infinite-dimensional CS become
inequivalent if applied to the finite-dimensional Hilbert spaces,
as will be shown in detail in this paper.

Optical Schr\"odinger cat states have attracted both
theoretical~\cite{Buzek95,SchleichBook} and
experimental~\cite{OpticalCats} interest in quantum optics,
quantum engineering and quantum information processing with
continuous variables.

Here we describe how to  generate superpositions of
macroscopically distinct QCS (i.e., Schr\"odinger cat-like states)
by the displacement of the vacuum in a Hilbert space for qudits.
We explain this counterintuitive result physically, in terms of
interference in phase space, and analytically by recalling the
properties of the roots of the Hermite polynomials. These cat
states are finite-dimensional analogs of the even and odd
infinite-dimensional CS. Some preliminary results, concerning the
generation of even QCS, were obtained in Ref.~\cite{Opatrny96}
(see also the review~\cite{Miran01}) but without a complete
analytical proof and a deeper physical explanation. Moreover, the
generation of odd QCS has not been predicted so far.

Wigner's~\cite{Wigner32} formulation of quantum mechanics based on
quasiprobabilities in phase space is completely equivalent to
other quantum formalisms including those of Schr\"odinger and
Heisenberg albeit \emph{without} the use of wave functions and
operators~\cite{Styer02}. The Wigner function is useful in quantum
optics~\cite{SchleichBook} in describing, e.g., interference in
phase space, and can be directly
measured~\cite{Lutterbach97,Bertet02} or indirectly reconstructed
using homodyne tomography both for infinite~\cite{SchleichBook}
and finite-dimensional~\cite{Vourdas04} systems.

In this paper, we apply the Wigner function formalism to study
both the interference in phase space and the homodyne detection of
QCS. Note that a nonstandard finite-dimensional Wigner
function~\cite{Wootters87,Leonhardt95} defined on a torus was
already applied to study some properties of QCS in
Refs.~\cite{Opatrny96,Miran01}. By contrast, here we apply the
standard Wigner function which, arguably, is simpler for physical
interpretations and for its measurement if the Hilbert-space
dimension is not very small or is unspecified.

Thus, we describe here nonclassical properties of the qudit cat
states revealed in their Wigner functions. In particular, we
analyze the nonclassical volume of the Wigner function, which is a
quantitative parameter of nonclassicality~\cite{Kenfack04}.

We also discuss optical tomograms, which can be directly measured
in the homodyne detection of the analyzed cat states.

This paper is organized as follows: In Secs.~II and~III, we
present two different constructions of QCS. We also describe their
Wigner representations and methods for their generation. In
Sec.~IV we show the main result of this paper: that the QCS
defined by the displacement of the vacuum can almost periodically
become the Schr\"odinger cat states having a clear interpretation
in terms of the Wigner function. We conclude in Sec.~V.

\section{Qudit coherent states by displacement of vacuum}

\subsection{Definition and Fock-state expansion}

In analogy to the first Glauber definition of the
infinite-dimensional CS, mentioned in the Introduction, one can
construct a QCS by applying a qudit displacement operator to the
vacuum~\cite{Buzek92}:
\begin{eqnarray}
\ket{\alpha}_d=\hat{D}_d(\alpha,\alpha^*)|0\rangle &=&
\exp(\alpha\hat{a}_d^{\dagger} -\alpha^* \hat{a}_d)|0\rangle,
\label{NCS}
\end{eqnarray}
where the qudit annihilation operator is defined by
\begin{eqnarray}
\hat{a}_d&=&\sum_{n=1}^{d-1}\sqrt{n} |n-1\rangle \langle n|
\label{N08}
\end{eqnarray}
and $\hat{a}_d^\dagger$ is the qudit creation operator, which are
the truncated versions of the usual infinite-dimensional
annihilation and creation operators, respectively. We note that
the commutator
\begin{eqnarray}
[\hat{a}_d,\hat{a}_d^\dagger]&=& 1-d \ket{d-1}\bra{d-1}
\label{N08a}
\end{eqnarray}
is a quantum number, which fundamentally differs from the
canonical commutation relation for the standard annihilation
$\hat{a}$ and creation $\hat{a}^\dagger$ operators. This
mathematical property implies that quantum interference in phase
space of the QCS is completely different from that of the standard
coherent states, as will be described in detail in the next
sections.

The Fock-state expansion of the QCS \NCS is much more complicated
than that for conventional CS and given by~\cite{Miran94}:
\begin{equation}
\ket{\alpha}_{d}=\sum_{n=0}^{d-1}c_{n}^{(d)}(\alpha)|n\rangle
\label{NQCS}
\end{equation}
with the superposition coefficients
\begin{eqnarray}
c_{n}^{(d)}(\alpha) & =f_{n}^{(d)} & \sum_{k=0}^{d-1}\frac{{\rm
He}_{n}(x_{k})}{[{\rm
He}_{d-1}(x_{k})]^{2}}\exp(ix_{k}|\alpha|),\label{c_n}
\end{eqnarray}
where
\begin{equation}
f_{n}^{(d)}=\frac{(d-1)!}{d}(n!)^{-1/2}
\exp[in\left(\phi_{0}-\tfrac{\pi}{2}\right)] \label{eq:f}
\end{equation}
and ${\rm He}_n(x)$ is the modified Hermite polynomial simply
related to the standard Hermite polynomial $H_n(x)$ as
\begin{equation}
{\rm He}_n(x)=2^{-n/2}H_n\left( x/{\sqrt{2}} \right).
\end{equation}
Moreover, $x_{k}\equiv x_{k}^{(d)}$ is the $k$th root of ${\rm
He}_{d}(x)$, and $\phi_0=\Arg(\alpha)$. For $d=2$, the general
formula, given by Eqs.~(\ref{NQCS}) and~(\ref{c_n}), reduces to
\begin{equation}
\ket{\alpha}_{2}  =  \cos(|\alpha|)|0\rangle+e^{
i\phi_{0}}\sin(|\alpha|)|1\rangle,\label{d2}
\end{equation}
which shows that any single-qubit pure state can be considered
this QCS for a proper choice of $\alpha$. Of course, this is not
the case for dimensions $d>2$. Two nontrivial examples for the
qutrit CS $\ket{\alpha}_{3}$ and quartit CS $\ket{\alpha}_{4}$ are
given in the Appendix.

Any finite superposition of Fock states, thus in particular the
QCS \NCS, can be realized by various experimental methods and
systems (see, e.g., Ref.~\cite{JMO97}). Here we just mention the
experiments of Zeilinger's group~\cite{Reck94} using generalized
Mach-Zehnder interferometers in a triangular configuration shown
in Fig.~1(c), and those of the Martinis group~\cite{Hofheinz09}
using microwave resonators coupled to superconducting quantum
circuits~\cite{Liu04}. It is also worth noting a probabilistic
method proposed in Ref.~\cite{DAriano00}, which uses a cross-Kerr
medium coupled to a ring cavity to synthesize arbitrary
superpositions of Fock states. Unfortunately, this method is based
on \emph{probabilistic} projective measurements contrary to the
method described below.

Let us now briefly describe the completely different approach of
Ref.~\cite{Leonski97}, shown schematically in Fig.~1(a). This
method enables, in principle, a direct and \emph{deterministic}
dynamical generation of the QCS \NCS for any amplitude $\alpha$
and small dimensions $d$. This is achieved by optical-state
truncation of the incident classical field by the so-called
nonlinear quantum scissors device composed of a higher-order Kerr
medium (modeled as a $d$-photon anharmonic oscillator) in a cavity
pumped by a classical driving field~\cite{Leonski97}. For this
reason, the QCS \NCS is sometimes referred to as the
\emph{nonlinear} QCS.

This optical truncation in the special case for $d=2$ results in
the celebrated single-photon blockade~\cite{Leonski94,Imamoglu97},
which is an effect when a single photon in a cavity with a Kerr
nonlinearity blocks the excitation of more photons in the cavity
field. The Kerr nonlinearity (which is proportional to the
third-order nonlinear susceptibility) can be induced relatively
easily by a strong interaction between the cavity field and
natural or artificial qubit~\cite{Buluta11,You11,Xiang13}, which
might be a single trapped atom~\cite{Birnbaum05}, a quantum
dot~\cite{Faraon08}, or a superconducting artificial
atom~\cite{Hoffman11,Lang11}. The single-photon blockade has been
mainly studied in the systems of cavity quantum electrodynamics
(QED) including theoretical predictions (see, e.g.,
Ref.~\cite{Tian92}) and experimental
demonstrations~\cite{Birnbaum05,Faraon08}. Recently, impressive
experimental progress was also reported in circuit-QED
systems~\cite{Hoffman11,Lang11}. Note that the two- and
three-photon blockades can be, in principle, observed in these
systems where the single-photon blockade was measured, but with
the choice of different resonance conditions~\cite{Miran13}. Other
generalized blockade effects comprise two-mode optical state
truncation~\cite{Leonski04} and single-\emph{phonon}
blockade~\cite{Liu10,Didier11}.

\begin{figure}
\includegraphics[scale=0.3]{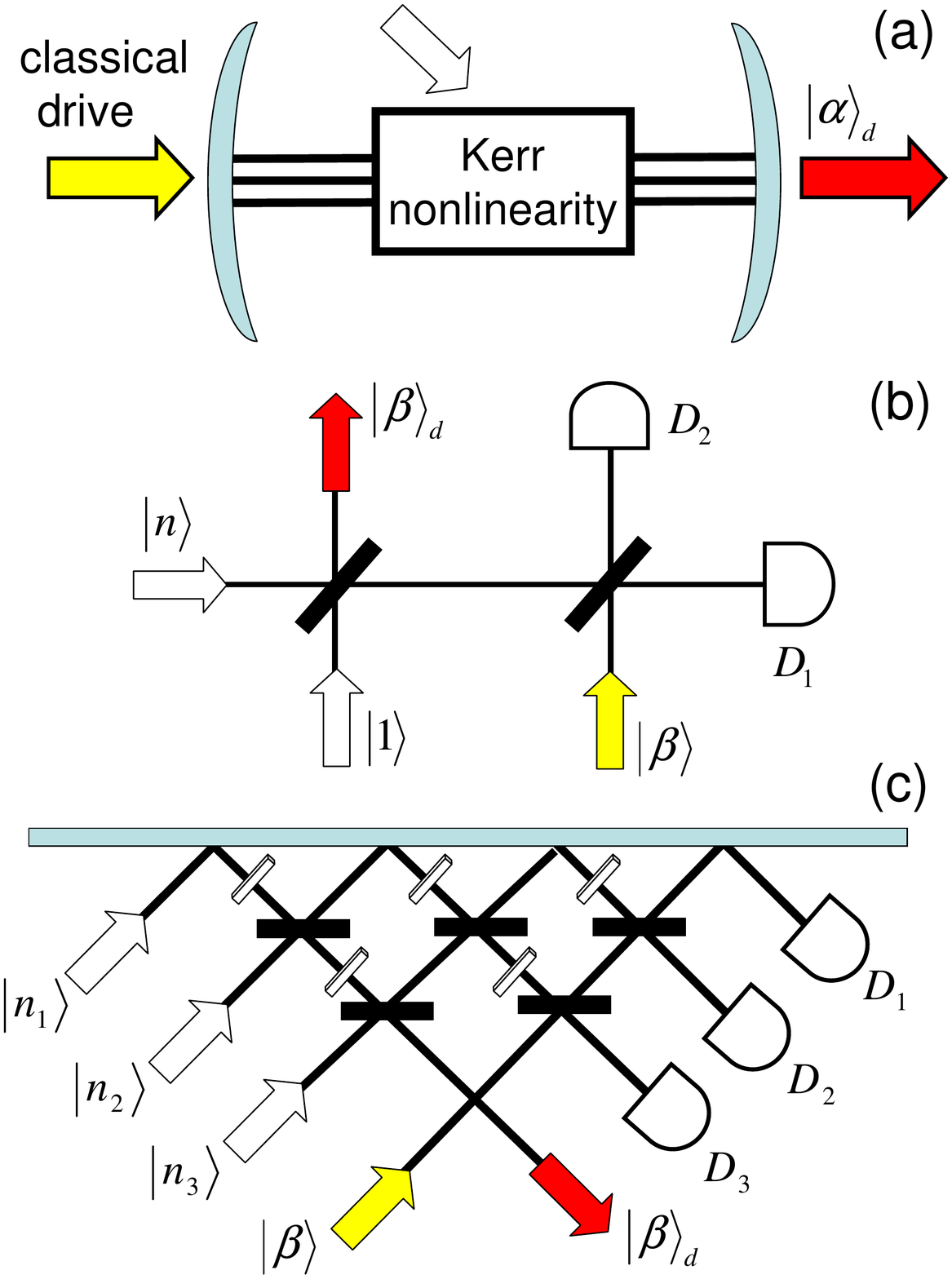}

\caption{(Color online) Examples of (a) nonlinear and (b,c) linear
quantum scissors devices for the generation of the QCS if losses
are negligible. Yellow (red) arrows denote input (output) fields,
white arrows are auxiliary fields, solid bars correspond to beam
splitters, blue bars to mirrors, empty bars to phase shifters, and
$D_n$ are photodetectors. (a) A cavity with a Kerr medium
($d$-photon anharmonic oscillator), described by a nonlinear
coupling proportional to the $(2d-1)$th-order nonlinear
susceptibility, driven by a classical laser light enables, in
principle, a deterministic generation of the QCS
$\ket{\alpha}_d$~\cite{Leonski97} (for $d=2$ see
Ref.~\cite{Imamoglu97} for details). (b) The Pegg-Phillips-Barnett
scissors~\cite{Pegg98} for the probabilistic generation (i.e.,
projection synthesis) of the QCS $\ket{\beta}_d$ with $d=2,3$ via
optical truncation and quantum teleportation of the incident CS
$\ket{\beta}$ conditioned on the proper photon-number detection at
the detectors $D_n$ using beam splitters with proper transmission
and reflection parameters, and the auxiliary Fock states $\ket{1}$
and $\ket{n}$ ($n=0$ for $d=2$ and $n=1$ for $d=3$). (c) A
generalized version of the Pegg-Phillips-Barnett scissors based on
a generalized Mach-Zehnder interferometer for a probabilistic
optical truncation and teleportation of $\ket{\beta}$ to the QCS
$\ket{\beta}_d$ with $d=2,...,6$~\cite{Miran05}. Note that the
configuration (c) is scalable for arbitrary $d$. It should be
stressed that the generation of QCS described here can be realized
also in various other bosonic finite-dimensional systems (see
text).}
\end{figure}

\subsection{Wigner representation of displaced vacuum for qudits}

The Wigner function associated with an arbitrary single-mode state
$\rho$ is defined by~\cite{Wigner32}
\begin{equation}
W(z)\equiv W_{\rho}(q,p)=\frac{1}{\pi}\int\bra{q-x}\rho \ket{q+x}
\exp\left(2ipx\right)dx, \label{wdef}
\end{equation}
where $q$ and $p$ are the canonical position and momentum
operators, respectively, and $z=q+ip$.

The concept of quantum interference in phase space for finite
superpositions of Fock states (so, in particular, for our QCS) can
be clearly explained in terms of the Wigner
function~\cite{Dowling91,Buzek95}, which will be discussed below.
Alternatively, one could explain this interference in a
semiclassical picture of the areas of overlap (i.e., interfering
areas)~\cite{Schleich87,SchleichBook}. Here we follow the
completely quantum approach of Ref.~\cite{Buzek95}.

The Wigner function for a qudit state, defined by
Eq.~(\ref{eq:state1}), can be given as a sum of two terms,
\begin{eqnarray}
  W(z) &=& W_{\rm mix}(z)+W_{\rm int}(z),
\label{W1}
\end{eqnarray}
representing, respectively, the noninterference (or mixture) part
for the Wigner function
\begin{eqnarray}
  W_{\rm mix}(z)&=& \sum_{n=0}^{d-1} |c_n|^2 W_n(z),
\label{W2}
\end{eqnarray}
which is given as a sum of the Wigner functions of the Fock states
$\ket{n}$,
\begin{eqnarray}
  W_{n}(z)&=& \frac2{\pi}(-1)^n \exp(-2|z|^2)L_n(4|z|^2),
\label{W3}
\end{eqnarray}
and the interference part
\begin{eqnarray}
  W_{\rm int}(z)&=& 2 \sum_{k<l}^{d-1} {\rm Re} [c^*_k c_l
  W_{kl}(z)],
\label{W4}
\end{eqnarray}
where
\begin{eqnarray}
  W_{kl}(z)= \frac2{\pi} (-1)^k
  \sqrt{\tfrac{k!}{l!}} (2z^*)^{l-k}  e^{-2|z|^2}L_k^{(l-k)}(4|z|^2),
\label{W5}
\end{eqnarray}
and $L_k^{(l-k)}(x)$ are the associated Laguerre polynomials
with $L_k(x)\equiv L_k^{(0)}(x)$. Equation~(\ref{W4}) can be
rewritten more compactly as
\begin{eqnarray}
  W_{\rm int}(z)&=& \frac4{\pi} e^{-2|z|^2}\sum_{k<l}
  (-1)^k |c_k| |c_l|  \sqrt{\tfrac{k!}{l!}}
  (2|z|)^{l-k} \nonumber\\ &&\hspace{2cm}\times L_k^{(l-k)}(4|z|^2)\cos(\Phi_{kl}),
\label{W6}
\end{eqnarray}
where $\Phi_{kl}=\Arg(c^*_k)+\Arg(c_l)+(k-l)\Arg(z)$.

It is seen that $W_{\rm mix}(z)$ and $W_{\rm int}(z)$ correspond,
respectively, to the diagonal and off-diagonal terms of the
density matrix $\rho=\ket{\psi}\bra{\psi}$ in Fock basis. The
Wigner function is \emph{phase insensitive} for Fock states
$\ket{n}$ (for any $n$) and their mixtures, so $W_{\rm mix}(z)$ is
symmetric for any rotations around $z=0$. By contrast, a
superposition of Fock states can be \emph{phase sensitive} as
described by the interference part $W_{\rm int}(z)$ of the Wigner
function, which explicitly depends on the phases $\Phi_{kl}$ (for
$k\neq l$), although the corresponding component Fock states
$\ket{k}$ and $\ket{l}$ are phase insensitive. Thus, interference
of probability amplitudes associated with off-diagonal terms of a
density matrix can be clearly described via interference in phase
space, although the Wigner-function approach is based on
probabilities (or rather quasiprobabilities as they  can be
negative) instead of probability amplitudes.

A few examples of the Wigner function for the qubit CS
$\ket{\alpha}_2$ are shown in Fig.~2, which can be compared with
the corresponding Wigner functions, shown in Fig.~3, for another
type of the qubit CS defined below. The Wigner functions for the
qutrit CS $\ket{\alpha}_3$ are shown in Fig.~4.  For clarity, we
rescaled colors in the plots of the Wigner functions such that
dark blue (dark red) corresponds to the minimum (maximum) values
in each figure. Blue regions correspond to the negative values of
the Wigner functions, which are the indicators of nonclassicality
of the states. Moreover, the black outer circles in these figures
show the areas in phase space, which are dominantly occupied by
the Wigner function for a given qudit state. Tails of the Wigner
function outside such circles can be practically ignored. Strictly
speaking, the Wigner function occupies the whole phase space for
any state, including the vacuum. But, the area where the Wigner
function is greater than an arbitrary threshold value is finite.
This area of phase space can be chosen arbitrarily. For example,
to describe an arbitrary $d$-dimensional state, we chose an area
large enough to cover the peak of the Wigner of an
infinite-dimensional CS $\ket{\alpha}$ with $|\alpha|^2=d-1$,
which is the photon number of the highest-energy Fock state in
${\cal H} ^{(d)}$, and the radius $r_0$ corresponding to its half
width at half maximum. This $r_0$ for a Gaussian curve with the
standard deviation $\sigma$ is equal to $r_0=\sigma\sqrt{2\ln 2}$.
The Wigner function for the CS $\ket{\alpha}$ is
$W_{\alpha}(q,p)=(2/\pi)\exp(-2|q+ip-\alpha|^2)$. Thus, the radius
of the outer circles in Figs.~2--6 was chosen as
\begin{equation}
r=\sqrt{d-1}+\sqrt{\ln 2/2}. \label{radius}
\end{equation}
By comparison, the inner circles in the plots of the Wigner
functions have the radius given by $r=\sqrt{d-1}$.

It is seen that the QCS \NCS with increasing $\alpha$
(corresponding to evolution time) is reflected from the boundary
states $\ket{0}$ and $\ket{d-1}$ of the Hilbert space ${\cal
H}^{(d)}$. This phenomenon of multiple reflections (multiple
bounce) can be interpreted as a ping-pong effect, which leads, in
particular, to the generation of the Schr\"odinger cat states as
will be shown in Sec.~IV.

\begin{figure}
\fig{\includegraphics[scale=0.35]{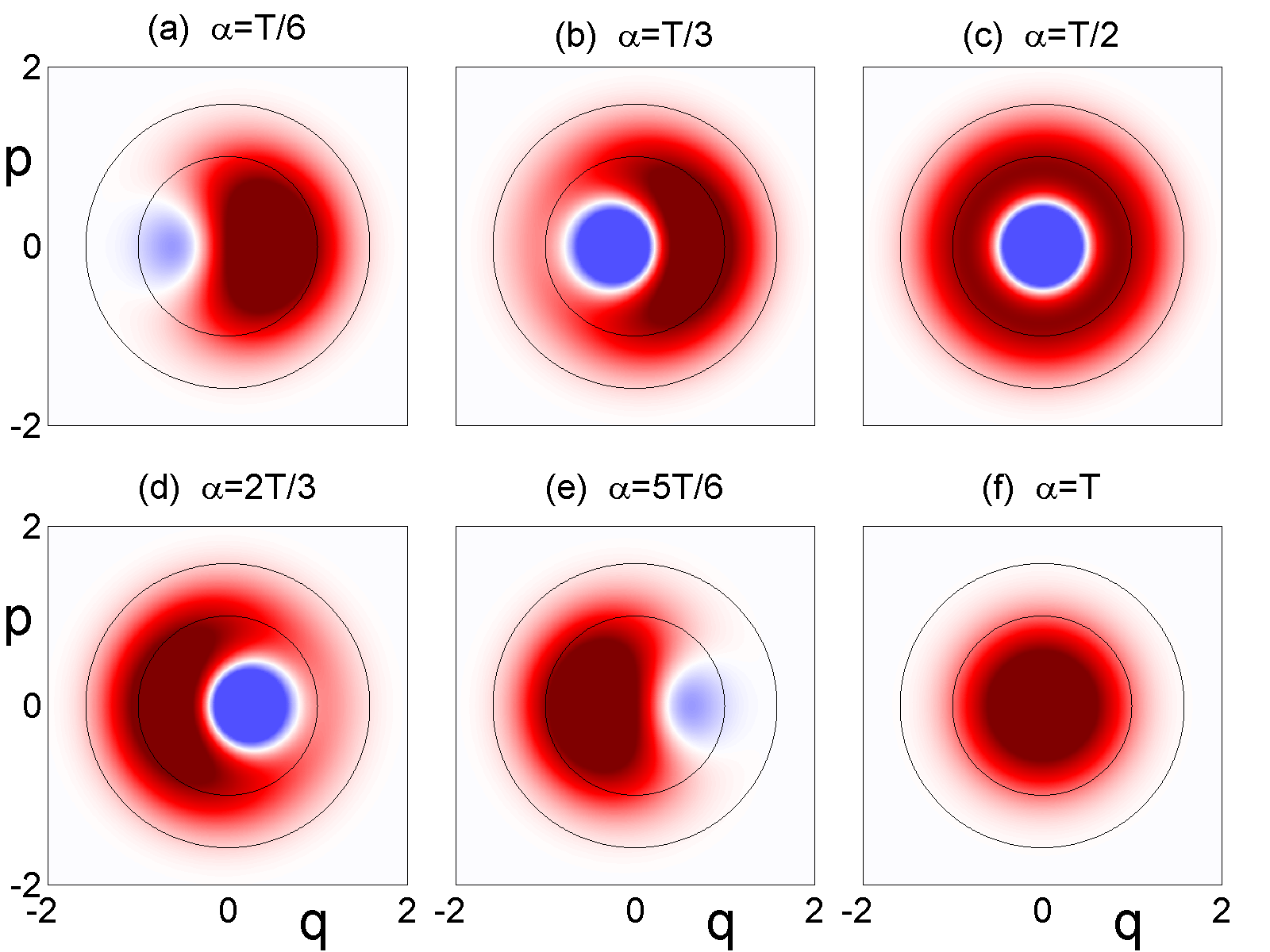}}

\caption{(Color online) Wigner functions for the \emph{qubit} CS
$\ket{\alpha}_2$ and various values of the real amplitude
$\alpha=n T/6$, where $T=T_2=\pi$. Note the snapshots of the
oscillations and the interference fringes in the Wigner function.
The increase of $\alpha$ can be interpreted as the evolution of
the driven Kerr system, shown in Fig.~1(a), assuming negligible
dissipation. The negative (positive) regions of the Wigner
function are marked in blue (red), with the deeper color the more
extreme values. Zero corresponds to white color. The inner and
outer circles have radii given by $r=\sqrt{d-1}$ and
Eq.~(\ref{radius}), respectively. It is seen that the Wigner
functions are practically vanishing beyond the outer circles.
Panels (c) and (f) show the Wigner functions for the single-photon
and vacuum states, respectively. The Wigner functions shown in
panels (a,b,d,e) are phase sensitive, which is a result of quantum
interference in phase space.}
\end{figure}
\begin{figure}
\fig{\includegraphics[scale=0.35]{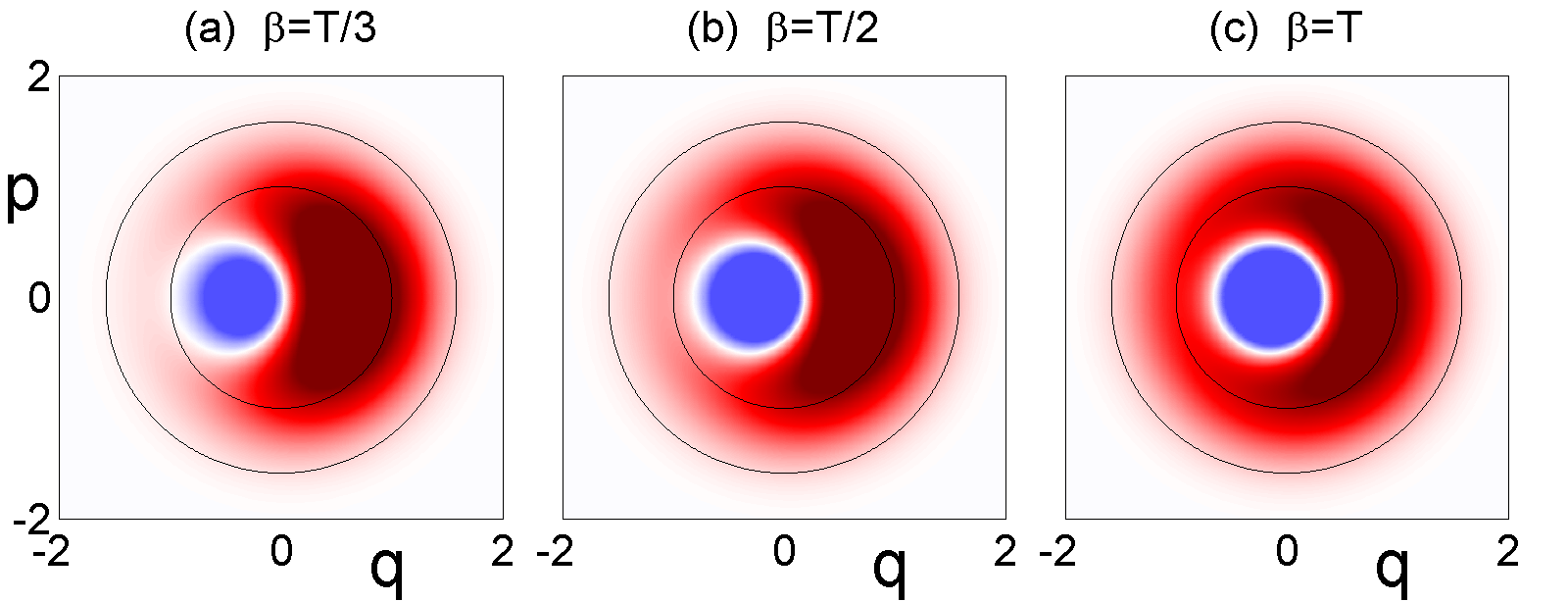}}

\caption{(Color online) As in Figs.~2(b,c,f) but for the
\emph{qubit} CS $|\beta\rangle_2$. Note that the Wigner function
for $|\beta=T/6\rangle_2$ resembles that for
$|\alpha=T/6\rangle_2$, as shown in Fig.~2(a), while for
$|\beta=2T/3\rangle_2$ and $|\beta=5T/6\rangle_2$ interpolates
between those in panels (b) and (c), where $T=T_2$. For brevity,
these three figures, corresponding to the cases shown in
Figs.~2(a,d,e), are omitted. In the limit of
$\beta\rightarrow\infty$, the state $|\beta\rangle_2$ goes into
the single-photon Fock state described by the standard
rotationally invariant Wigner function.}
\end{figure}
\begin{figure}
\fig{ \includegraphics[scale=0.35]{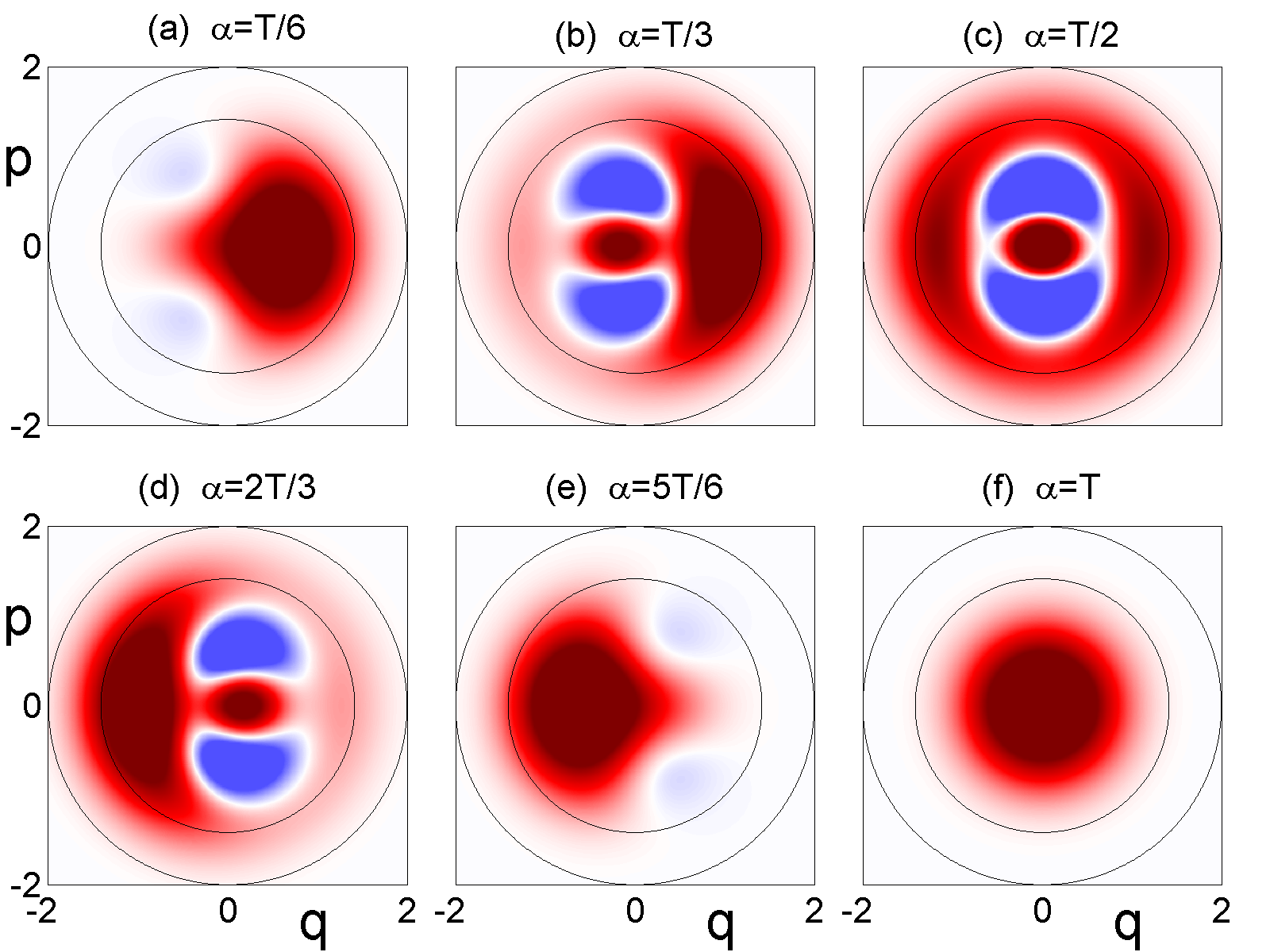}}

\caption{(Color online) Wigner functions for the \emph{qutrit}
($d=3$) CS $\ket{\alpha}_3$ with $\alpha=n T_3/6$ with
$n=$1,2...\,. The color codes and circles correspond to those in
Fig.~2. Panels (c) and (f) show the Wigner functions for a cat
state (even QCS) and the vacuum, respectively.}
\end{figure}
\begin{figure}
\fig{\includegraphics[scale=0.35]{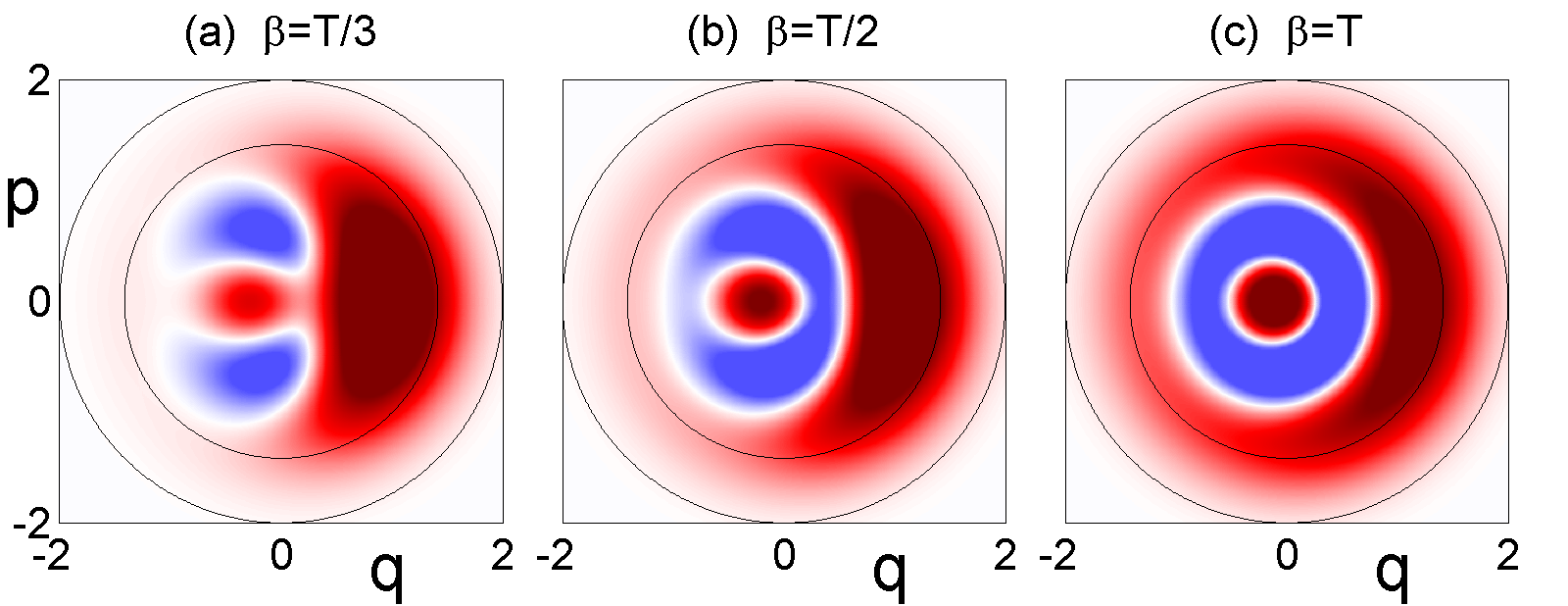}}

\caption{(Color online) Same as in Figs.~4(b,c,f) but for the
\emph{qutrit} CS $|\beta\rangle_3$. Analogously to Fig.~3, the
Wigner function for $|\beta=T/6\rangle_3$ with $T=T_3$ resembles
that for $|\alpha=T/6\rangle_3$, as shown in Fig.~4(a), while for
$|\beta=2T/3\rangle_3$ and $|\beta=5T/6\rangle_3$ interpolates
between those in panels (b) and (c). For brevity, these three
figures, corresponding to Figs.~4(a,d,e), are not presented here.
Note that the limiting state
$\lim_{\beta\rightarrow\infty}|\beta\rangle_3=|2\rangle$ is
described by the standard rotationally invariant Wigner function
of the two-photon Fock state.}
\end{figure}

\section{Qudit coherent states by truncation of Fock-state expansion of Glauber coherent states}

Another type of the QCS can be simply obtained by truncating the
Fock-state superposition of the conventional infinite-dimensional
CS as was studied by, e.g., Kuang \etal~\cite{Kuang93}. To be
precise, this QCS can defined by
\begin{equation}
|\beta\rangle_{d}= {\cal N}
\exp(\beta\hat{a}_d^{\dagger})|0\rangle = {\cal
N}\sum_{n=0}^{d-1}\frac{\beta^{n}}{\sqrt{n!}}|n\rangle\label{LCS}
\end{equation}
for a complex amplitude $\beta$. This definition is postulated in
analogy to the second Glauber definition of the conventional CS
based on the Campbell-Baker-Hausdorff theorem as follows
\begin{equation}
e^{\hat{A}+\hat{B}}\ket{0}=e^{\hat{A}} e^{\hat{B}}
e^{\hat{C}}\ket{0}=e^{\hat{C}}e^{\hat{A}}\ket{0}={\cal
N}\exp(\beta \hat{a}^\dagger)\ket{0} \label{baker}
\end{equation}
where $\hat{A}=\hat{B}^\dagger=\beta \hat{a}^\dagger$,
$\hat{C}=-\frac12 [\hat{A},\hat{B}]$, and ${\cal
N}=e^{\hat{C}}=\exp{(-\frac12 |\beta|^2)}$. This theorem can be
applied to the infinite-dimensional operators since it holds
$[\hat{A},[\hat{A},\hat{B}]]=[\hat{B},[\hat{A},\hat{B}]]=0$. By
contrast, the Campbell-Baker-Hausdorff theorem cannot be applied
to the finite-dimensional annihilation and creation operators
since the double commutators
$[\hat{a}_d,[\hat{a}_d,\hat{a}_d^\dagger]]$ and
$[\hat{a}^\dagger_d,[\hat{a}_d,\hat{a}_d^\dagger]]$ do not vanish,
as can be seen by applying Eq.~(\ref{N08a}). Thus, the two kinds
of QCS, as defined by Eqs.~(\ref{NCS}) and~(\ref{LCS}), are
fundamentally different (except some special cases)  exhibiting
different quantum interference in phase space, as seen in
Figs.~2--6.

One can refer to $|\beta\rangle_{d}$ as the \emph{linear} CS for a
qudit since it can be simply (but non deterministically) obtained
by linear optical systems called linear quantum scissors, as shown
schematically in Figs.~1(b) and~1(c) and described in detail for
$d=2$ in Refs.~\cite{Pegg98,Ozdemir01}, $d=3$~\cite{Koniorczyk00},
and higher $d$~\cite{Miran05}. For $d=2$, Eq.~(\ref{LCS}) reduces
to the qubit CS $|\beta\rangle_{2}= {\cal
N}(\ket{0}+\beta\ket{1})$. Although the systems shown in
Figs.~1(b,c) seemingly contain only linear optical elements, the
\emph{nonlinearity} is induced by the measurement (i.e., the
conditional photodetection). So, the generation of the QCS \LCS
also requires nonlinearity. Nevertheless, the term \emph{linear}
QCS stresses only the fact that no nonlinear media are used in the
setups of Figs.~1(b,c).

The Wigner functions for $|\beta\rangle_{2}$ are shown in Fig.~3,
which could be compared with those for $|\alpha\rangle_{2}$ in
Fig.~2 for some particular choices of $\alpha=\beta$. Analogously,
Figs.~4 and~5 of the Wigner functions for the qutrit CS
$|\alpha\rangle_{3}$ and $|\beta\rangle_{3}$, respectively, show
similar properties of the states for $|\alpha|=|\beta|\ll T_3/2$
[in Figs.~4(b) and~5(a)] and their distinctive properties for
other values of $|\alpha|=|\beta|$ [in Figs.~4(c,f) and~5(b,c)].

It is seen that the QCS \LCS is not reflected from the boundaries
of the Hilbert space as $\beta$ increases. This can be described
as ``no bouncing''. By contrast, as already mentioned, the QCS
\NCS exhibits multiple bounce (or a ping-pong effect) as $\alpha$
increases.

One can define a state $\ket{\gamma}_d$ complementary to the QCS
$\ket{\beta}_d$, such that their equally weighted superposition is
the QCS:
\begin{equation}
  \ket{\alpha}_d={\cal N} (\ket{\beta}_d+\ket{\gamma}_d),
  \label{gamma1}
\end{equation}
which leads to the explicit form of the complementary state
\begin{equation}
  \ket{\gamma}_d=2\, _d\langle\alpha|\beta\rangle_d \,\ket{\alpha}_d-\ket{\beta}_d
  \label{gamma}
\end{equation}
up to a global phase factor. In the simplest case for $d=2$, one can find
\begin{eqnarray}
  \ket{\gamma}_2 &=& \frac{1}{\sqrt{1+\alpha^{2}}}
  \Big([\cos(2\alpha)+\alpha\sin(2\alpha)]\ket{0}
  \nonumber \\ &&+ [\sin(2\alpha)-\alpha\cos(2\alpha)]\ket{1} \Big),
\label{N1}
\end{eqnarray}
where for simplicity we assumed $\alpha$ to be positive. By
contrast, the qubit CS $\ket{\beta}_2$ is given by ${\cal N}
(\ket{0}+\beta\ket{1})$, as depicted for some choices of $\beta$
in Fig.~3. Thus, for the choice of $\alpha=\beta=\gamma=T_2/2$, we
have  $\ket{\gamma}_2 = {\cal N}
(-\ket{0}+\frac{\pi}2\ket{1})=-\ket{-\beta}_2$, which results in
$\ket{\alpha}_2=\ket{1}$. We note that such a simple relation
between $\ket{\gamma}_d$ and $\ket{-\beta}_d$ exists for $d=2$
only. An explicit comparison of $\ket{\gamma}_d$ and
$\ket{-\beta}_d$ for $d=3,4$ is given in the Appendix.

\begin{figure}
\fig{ \includegraphics[scale=0.35]{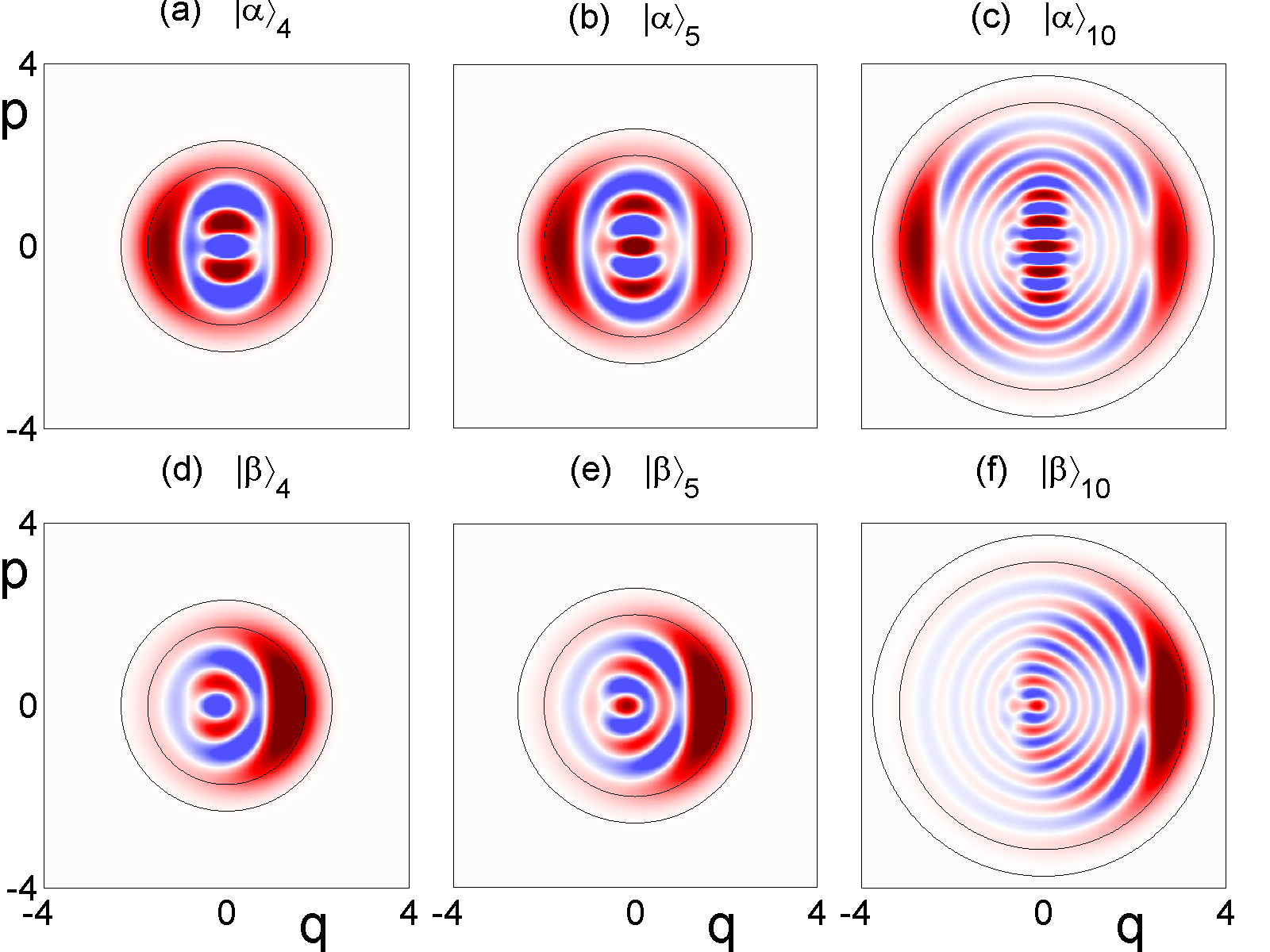}}

\caption{(Color online) Wigner functions for the \emph{qudit} CS
$\ket{\alpha}_d$ (a,b,c) and $|\beta\rangle_d$ (d,e,f) with
$\alpha=\beta=T_d/2$ for $d=4,5,10$, respectively. The
corresponding plots for $d=2,3$ are shown in Figs.~2--5 for
$\alpha=\beta=T_d/2$. The color codes are the same as in Fig.~2.
Panels (a,c) and (b) show the Wigner functions for cat states: the
odd ($\ket{\alpha_{-}}_d$) and even ($\ket{\alpha_{+}}_d$) QCS,
respectively.  We note that $\ket{\alpha_{\pm}}_d$ are also very
close to $\ket{\beta_{\pm}}_d$ as revealed by their fidelities
close to 1, which are shown in Table~I. All these Wigner functions
are phase sensitive due to quantum interference in phase space.}
\end{figure}

\begin{figure}
\fig{
\includegraphics[scale=0.31]{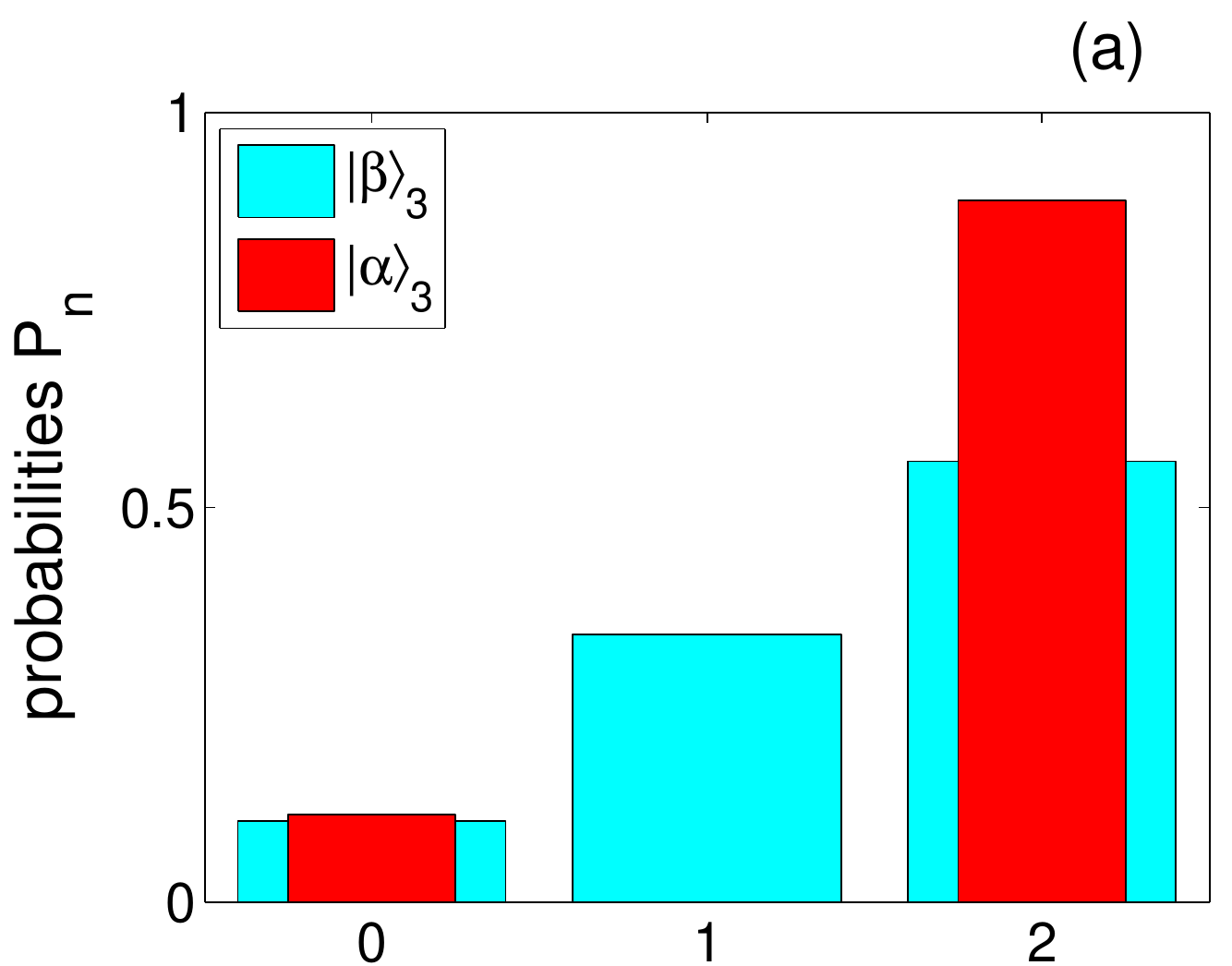}
 \includegraphics[scale=0.31]{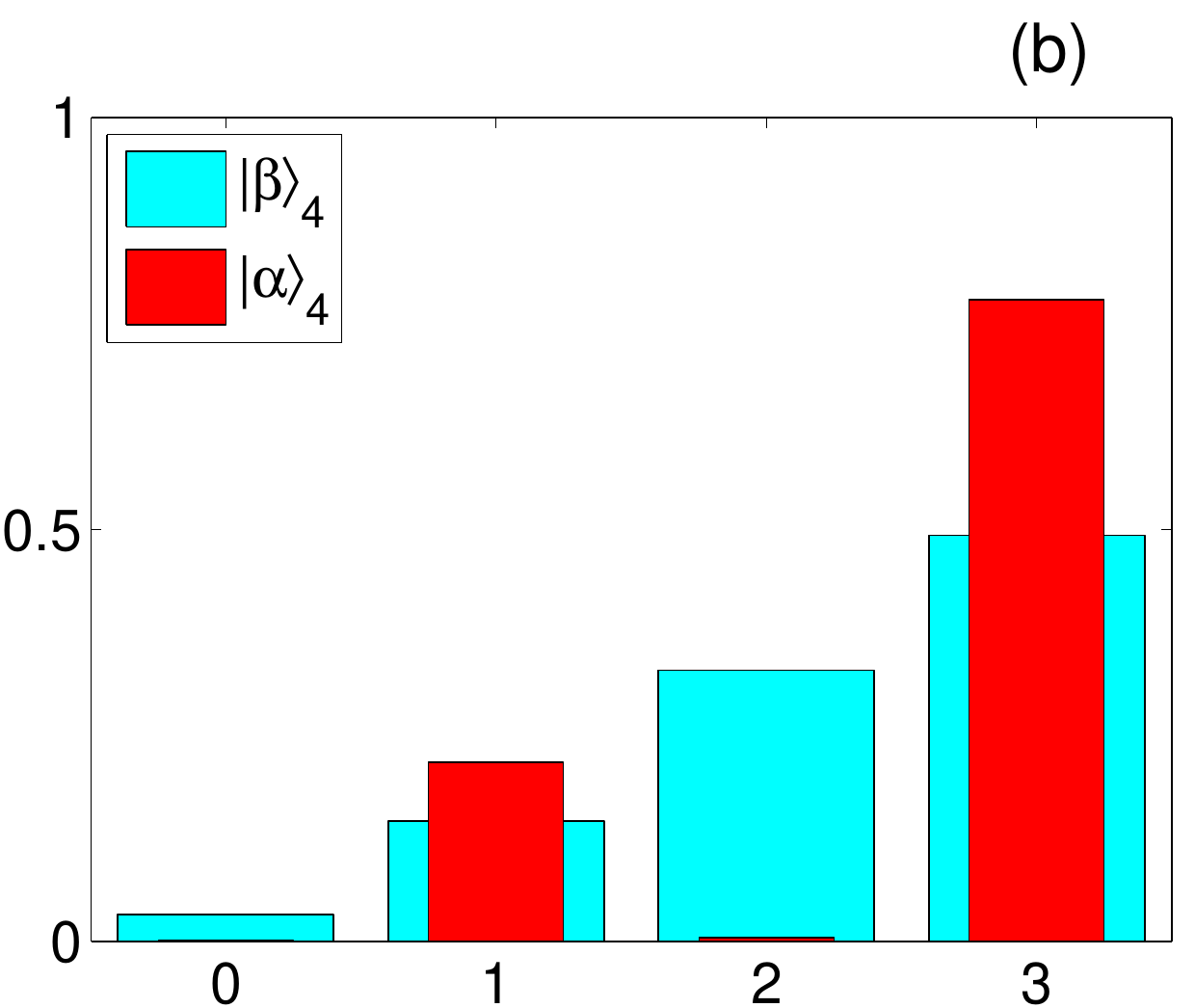}}

\fig{
  \includegraphics[scale=0.31]{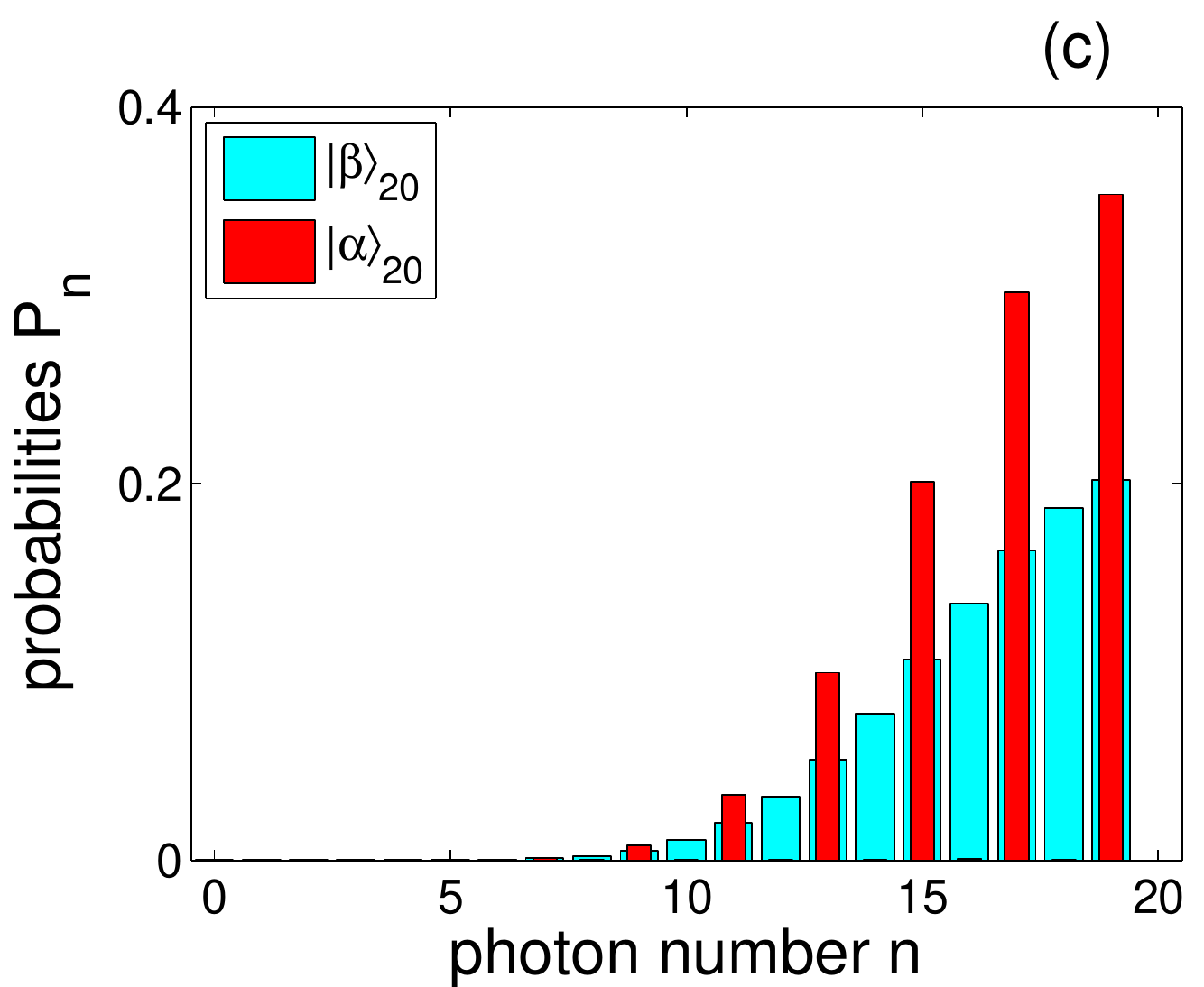}
   \includegraphics[scale=0.31]{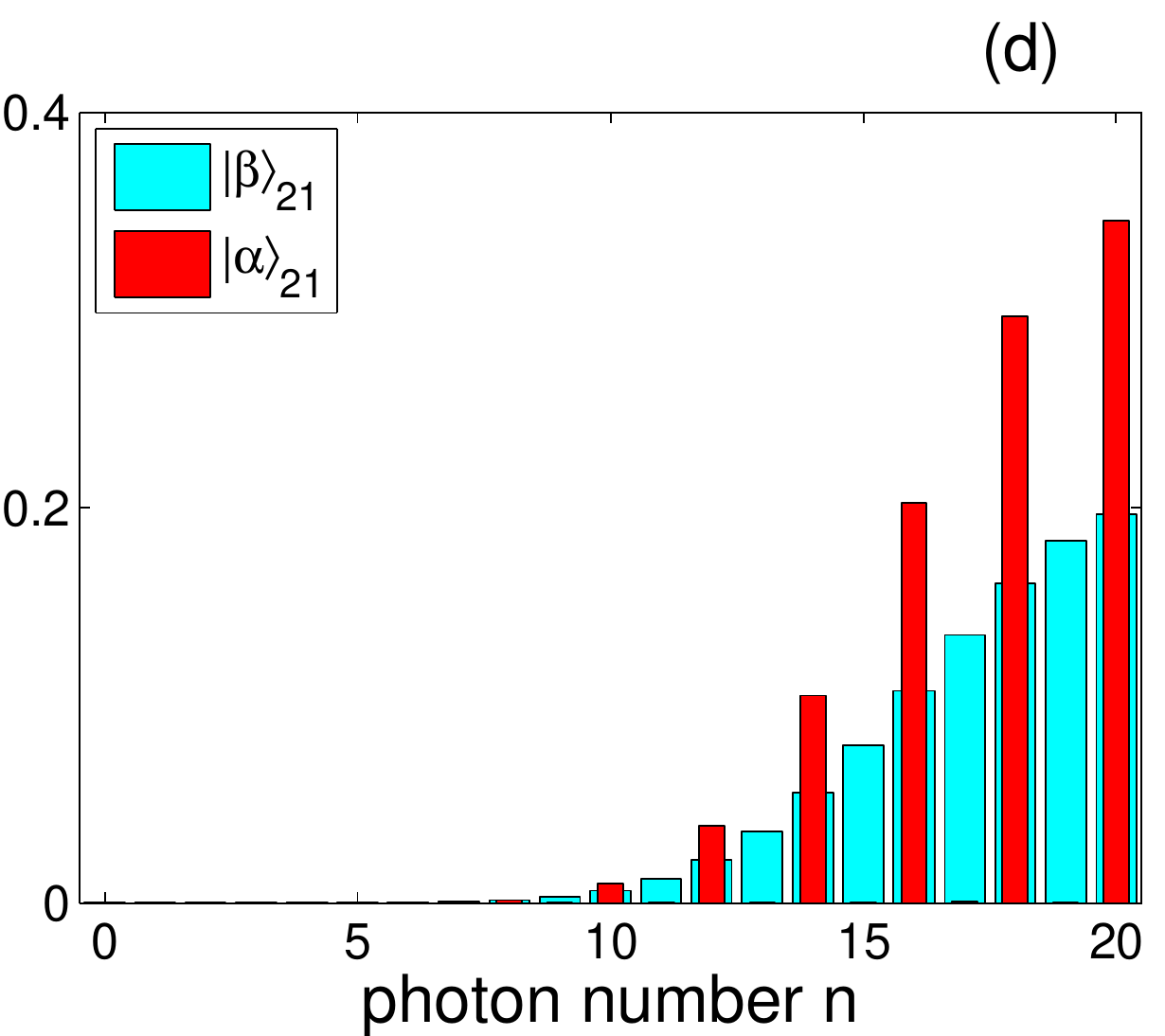}}

\caption{(Color online) Photon-number distributions
$P_n(\alpha)=|\langle n\ket{\alpha}_{d}|^2$ (red thin) and
$P_n(\beta)=|\langle n|\beta\rangle_{d}|^2$ (broad cyan bars) for
the QCS $\ket{\alpha}_{d}$ and $\ket{\beta}_{d}$ with
$\alpha=\beta=T_d/2$ and various $d$. It is seen  that the
Schr\"odinger cat states are generated: the even QCS
$\ket{\alpha}_{d}=\ket{\alpha_+}_d$ for (a) $d=3$ and (d) $d=21$,
while the odd QCS $\ket{\alpha}_{d}=\ket{\alpha_-}_d$  for (b)
$d=4$ and (c) $d=20$.}
\end{figure}

\section{Cat-state generation}

Here we will show one of the main results of this paper: that
macroscopically distinguishable superpositions of the QCS
(Schr\"odinger cat states) can be simply generated by displacing
the vacuum in the Hilbert space of an optical qudit.

\subsection{Even and odd coherent states for qudits}

The prototype examples of optical Schr\"odinger cat states are the
even and odd infinite-dimensional CS, defined~\cite{Buzek95} as
$|\alpha_{\pm}\rangle = {\cal N} ( |\alpha \rangle \pm |-\alpha
\rangle)$, being also referred to as the Schr\"odinger male and
female cat states, respectively.

By analogy with the infinite-dimensional cat states
$|\alpha_{\pm}\rangle$, one can define their qudit counterparts
as, e.g.,  the even QCS, $|\alpha_{+}\rangle_d$, and odd QCS,
$|\alpha_{-}\rangle_d$, as follows:
\begin{eqnarray}
|\alpha_{+}\rangle_d = {\cal N}( |\alpha \rangle_d
+|-\alpha\rangle_d) ={\cal N}
\sum_{n=0}^{d-1}\frac{c_{2n}^{(d)}(\alpha)|2n\rangle}{\sqrt{(2n)!} } ,\hspace{9mm}\label{N12} \\
|\alpha_{-}\rangle_d\! =\! {\cal N} ( |\alpha \rangle_d -|-\alpha
\rangle_d ) ={\cal N}
\sum_{n=0}^{d-1}\frac{c_{2n+1}^{(d)}(\alpha)|2n+1\rangle}{\sqrt{(2n+1)!}
},\;\; \label{N13}
\end{eqnarray}
where the superposition coefficients $c_{n}^{(d)}(\alpha)$
are given by Eq.~(\ref{c_n}). Moreover, one
can define other qudit cat states based on the QCS $|\pm\beta
\rangle_{d}$ as follows:
\begin{eqnarray}
|\beta_{+}\rangle_{d} = {\cal N}( |\beta \rangle_{d} +|-\beta \rangle_{d} )
={\cal N} \sum_{n=0}^{d-1 }\frac{\beta
^{2n}|2n\rangle}{\sqrt{(2n)!} } ,\hspace{9mm}\label{N10} \\
|\beta_{-}\rangle_{d} = {\cal N}( |\beta \rangle_{d} -|-\beta \rangle_{d} )
={\cal N} \sum_{n=0}^{d-1 }\frac{\beta
^{2n+1}|2n+1\rangle}{\sqrt{(2n+1)!} }. \label{N11}
\end{eqnarray}

In the following, we will explain why the QCS $\ket{\alpha}_d$ for
$\alpha=T_d/2$ are very good approximations of either the even QCS
$|\alpha_{+}\rangle_d$ and $|\beta_{+}\rangle_d$ for odd $d$ or
the odd QCS $|\alpha_{-}\rangle_d$ and $|\beta_{-}\rangle_d$ for
even $d$.

\subsection{Periodicity, quasiperiodicity and symmetries of Wigner functions}

As found in Refs.~\cite{Opatrny96,Leonski97b}, the QCS \NCS with
increasing $\alpha$ exhibit either perfect periodicity for $d=2,3$
or almost periodicity (``quasiperiodicity'') for $d>3$. The
periods for $d=2,3$ are $T_2=\pi$ and $T_3=2\pi/\sqrt{3}$,
respectively, while the quasiperiod $T_d$ for $d>3$ is given by:
\begin{equation}
T_d=\sqrt{4d+2}. \label{period}
\end{equation}
Note that Eq.~(\ref{period}) gives a rough approximation even for
$T_2$ (as $\pi = \sqrt{10}-0.02...$) and $T_3$ (as $2\pi/\sqrt{3}
= \sqrt{14}-0.1...$).

The period $T_2$ is equal to $\pi$ up to a global phase since
$\ket{\alpha}_{2}=-|\alpha+\pi\rangle_{2}$ [compare the Wigner
function for $|\alpha=\pi\rangle$ in Fig.~2(f) , which is the same
as for $|\alpha=0\rangle$]. Obviously, by doubling the period,
this extra global phase does not appear. It was discussed in
Ref.~\cite{Opatrny96} that the quasiperiod $T_d$ of even $d$ is
twice larger than that for odd $d$. Nevertheless if one ignores
the global $\pi$-shift  (which is usually physically justified)
then Eq.~(\ref{period}) determines the quasiperiods of the QCS
\NCS both for the even and odd dimensions $d$.

By analyzing Figs.~2 and~4, one can find that
$W(q,p;|T_d-\alpha\rangle_{d})$ is just $W(q,p;\ket{\alpha}_{d})$
but rotated by $\pi$ in phase space. This can be easily understood
by recalling the exact symmetries for $d=2,3$:
\begin{equation}
  W(q,p;|T_d-\alpha\rangle_{d})=W(q,p;|-\alpha\rangle_{d})=W(-q,-p;\ket{\alpha}_{d}).
  \label{WignerSymmetry}
\end{equation}
Analogous approximate symmetries hold for the quasiperiods $T_d$
with $d>3$. These properties imply that $W(q,p;|T_d/2\rangle_{d})$
are perfectly symmetric [as shown in Figs.~2(c) and 4(c)] or
approximately symmetric [see Figs.~6(a,b,c)] along the line $q=0$
in phase space.

The state $\ket{\alpha}_{2}$ for $\alpha=T_2/2$ [as shown in
Fig.~2(c)] is just a single-photon Fock state, so it can hardly be
considered a real cat state. The simplest nontrivial cat state
$|\alpha=T_d/2\rangle_{d}$, as a superposition of the two
out-of-phase QCS, exists for $d=3$ as given by
\begin{eqnarray}
|\alpha=\tfrac12 T_3\rangle_{3} & = &
\frac{1}{3}(|0\rangle+2\sqrt{2}e^{ i2\phi_{0}}|2\rangle),
\label{cat3}
\end{eqnarray}
which follows from Eq.~(\ref{d3}).

\subsection{Analytical explanation of cat-state generation by displacing vacuum}

Here we show that \NCS quasiperiodically evolves into the odd
(even) QCS for an even (odd) dimension $d>3$. In addition, the
\emph{exact} periodic generation of the even QCS
$\ket{\alpha_+}_3$ for $d=3$ is shown explicitly in the Appendix.

First, by recalling the reflection formula ${\rm
He}_{n}(-x)=(-1)^n {\rm He}_{n}(x),$ we find that Eq.~(\ref{c_n})
for even $n$ (and any $d$) can be rewritten as
\begin{eqnarray}
c_{n}^{(d)}(\alpha) & = &
2f_{n}^{(d)}\sum_{l=1}^{\sigma}\frac{{\rm
He}_{n}\left(x_{l}\right)}{[{\rm
He}_{d-1}\left(x_{l}\right)]^{2}}\cos(x_{l}|\alpha|) \nonumber\\
&&+\delta_{d,{\rm odd}}f_{n}^{(d)}\frac{{\rm He}_{n}(0)}{[{\rm
He}_{d-1}(0)]^{2}},\label{c_n_even}
\end{eqnarray}
while for odd $n$ as
\begin{eqnarray}
c_{n}^{(d)}(\alpha) & = &
2if_{n}^{(d)}\sum_{l=1}^{\sigma}\frac{{\rm
He}_{n}\left(x_{l}\right)}{[{\rm
He}_{d-1}\left(x_{l}\right)]^{2}}\sin(x_{l}|\alpha|),\label{c_n_odd}
\end{eqnarray}
where $\sigma={\rm int}(d/2)$ is the integer part of $d/2$,
$f_{n}^{(d)}$ is defined by Eq.~(\ref{eq:f}), and $x_l$ for
$l=1,...,\sigma$ denote only positive roots of ${\rm He}_{d}(x)$,
contrary to $x_k$ in Eq.~(\ref{c_n}) corresponding to all $d$
roots. The Hermite polynomials in the last term in
Eq.~(\ref{c_n_even}) can be explicitly given in terms of the Euler
$\Gamma$ function as ${\rm He}_{n}(0)=\sqrt{\pi 2^{n}}/
\Gamma[(1-n)/2]$.

Then, we apply oscillatory functions approximating well the
Hermite polynomials for small $|x|\ll \sqrt{2n}$, which can be
given for even $n$ as follows~\cite{SpanierBook}:
\begin{equation}
 {\rm He}_{n}(x)\approx i^{n}(n-1)!!\exp(\tfrac14 x^{2})
 \cos\left(x\sqrt{n+\tfrac12}\right)
 \label{He_even}
\end{equation}
and for odd $n$ as
\begin{equation}
  {\rm He}_{n}(x)\approx-i^{n+1}\frac{n!!}{\sqrt{n}}
  \exp(\tfrac14x^{2})\sin\left(x\sqrt{n+\tfrac12}\right).
 \label{He_odd}
\end{equation}
Thus, it is readily seen from Eqs.~(\ref{He_even})
and~(\ref{He_odd}) that the roots of ${\rm He}_{d}(x)$ for
$l=-(d-1),-(d-3),...,(d-3),(d-1)$ are
\begin{equation}
x_{l}^{(d)}\approx \frac{l\pi}{\sqrt{4d+2}},\label{roots}
\end{equation}
which results in Eq.~(\ref{period}) for the quasiperiod $T_d$ of
\NCS if the global phase of \NCS is ignored. Note that in
Eqs.~(\ref{c_n_even}) and~(\ref{c_n_odd}), the roots $x_{l}^{(d)}$
are considered for positive $l$ only. Equation~(\ref{roots}) also
implies that $x_l^{(d)}T_d/2\approx l\pi/2$. Thus, by applying
this result to Eqs.~(\ref{c_n_even}) and~(\ref{c_n_odd}), we have
(for $n=0,1,...$):
\begin{eqnarray}
  c_{2n}^{(2\sigma)}(\tfrac12 T_{2\sigma}) \approx 0, \quad
  c_{2n+1}^{(2\sigma)}(\tfrac12 T_{2\sigma}) \neq 0,
\label{c1}
\end{eqnarray}
corresponding to the generation of the odd QCS for an even
dimension $d=2\sigma$, and
\begin{eqnarray}
  c_{2n}^{(2\sigma+1)}(\tfrac12 T_{2\sigma+1}) \neq 0, \quad
  c_{2n+1}^{(2\sigma+1)}(\tfrac12 T_{2\sigma+1}) \approx 0,
\label{c2}
\end{eqnarray}
which explains the generation of the even QCS for an odd dimension
$d=2\sigma+1$. Finally, we can write
\begin{eqnarray}
  \ket{\alpha=\tfrac12 T_{2\sigma}}_{2\sigma}&\approx&\ket{\alpha_{-}}_{2\sigma}\approx\ket{\beta_{-}}_{2\sigma}, \label{c3}
\\
  \ket{\alpha=\tfrac12 T_{2\sigma+1}}_{2\sigma+1}&\approx&\ket{\alpha_{+}}_{2\sigma+1}\approx\ket{\beta_{+}}_{2\sigma+1}. \label{c4}
\end{eqnarray}
where the relations for $\ket{\beta_{\pm}}_d$ are given on the
basis of their definitions and our numerical calculations
discussed in the next section and summarized in Table~I.

\subsection{Photon-number distributions and fidelities of the cat-state generation}

Figure~7 shows the photon-number distributions for the QCS \NCS
and \LCS assuming $\alpha=\beta$ to be in the middle of the
quasiperiod $T_d$ for $d=3,4,20,21$. It is seen that every second
term in all these cases of \NCS is practically vanishing on the
scale of the figures. This is in contrast to the photon-number
distribution for \LCS, which is a truncated Poissonian
distribution of the conventional Glauber CS. Thus, Fig.~7 confirms
our predictions that \NCS corresponds either to even or odd QCS
depending on the parity of the dimension $d$. It is worth noting
that the photon-number oscillations in the QCS \NCS are a clear
signature of quantum interference in phase space. This can be
described even semiclassically in analogy to the explanation of
the photon-number oscillations for squeezed
states~\cite{Schleich87,SchleichBook}.

To show how well the QCS can approximate the cat states, we
calculate the fidelities between various states, as shown in
Table~I. As already mentioned, for $d=2$ and $\alpha=\beta=\pi/2$,
the qubit cat states are singular, because they correspond to a
single-photon Fock state, i.e.,
$\ket{\alpha}_2=\ket{\alpha_{-}}_2=\ket{\beta_{-}}_2=\ket{1}$ [as
shown in Figs.~2(c) and~8(a)], which results in the perfect
fidelities between these states. The lowest-dimensional nontrivial
QCS corresponding to a cat state can be observed for $d=3$ and
$\alpha=\beta=T_3/2$, as we have
$\ket{\alpha}_3=\ket{\alpha_{+}}_3$, given by Eq.~(\ref{cat3})
[see Figs.~4(c), 7(a) and 8(b)], which is similar but not exactly
equal to $\ket{\beta_{+}}_3$. These properties result in
$|_3\langle\alpha|\alpha_{+}\rangle_{3}|^2=1$ and
$|_3\langle\alpha|\beta_{+}\rangle_{3}|^2<1$. As already
mentioned, there is a perfect periodicity of $\ket{\alpha}_d$ as a
function of $\alpha$ for $d=2,3$, and only quasiperiodicity for
$d\ge 4$.

The lowest fidelities $\fidelity{\alpha}{\alpha_{\pm}}$ and
$\fidelity{\alpha}{\beta_{\pm}}$ among any dimension $d$ if
$\alpha=\beta=T_d/2$ are achieved for $d=4$ [see Figs.~6(a), 7(b)
and 8(c)] as the accuracy of the quasiperiod $T_4$ of
$\ket{\alpha}_4$ is the worst for this dimension among any finite
$d$. Nevertheless, this worst case still corresponds to the
relatively high fidelities, i.e.,
$\fidelityD{\alpha}{\alpha_{-}}{4}\approx\fidelityD{\alpha}{\beta_{-}}{4}\approx
0.995$. By contrast, the generated cat states \NCS for
$\alpha=T_d/2$ are clearly different from the mixed states
\begin{equation}
\rho^{(d)}_{\rm mix}=\tfrac 12 (\ket{\beta}_d\,
_d\bra{\beta}+\ket{-\beta}_d\, _d\bra{-\beta}) \label{rho_mix}
\end{equation}
with $\alpha=\beta$. This is shown in Table~I for the
fidelities
\begin{equation}
F^{(d)}_{\rm mix}=\,_d\bra{\alpha}\rho^{(d)}_{\rm
mix}\ket{\alpha}_d, \label{Fmix}
\end{equation}
which, together with $\fidelity{\alpha}{\beta}$, are evidently
much smaller than the other fidelities listed there.

\begin{table}
\caption{Comparison of the fidelities for the QCS $\ket{\alpha}_d$
and $\ket{\beta}_d$, and the corresponding cat states
$\ket{\alpha_{\pm}}_d$ and $\ket{\beta_{\pm}}_d$ assuming
$\alpha=\beta=T_d/2$ and the sign $+$ ($-$) is chosen for the odd
(even) $d$-dimensional Hilbert space. Additionally, $F^{(d)}_{\rm
mix}$ is given by Eq.~(\ref{Fmix}). }
\begin{tabular}{r c c c c c}
  \hline
      $d$ &
      $\fidelity{\alpha}{\beta}$ &
      $\fidelity{\alpha}{\alpha_{\pm}}$ &
      $\fidelity{\alpha}{\beta_{\pm}}$ &
      $\fidelity{\alpha_{\pm}}{\beta_{\pm}}$ &
      $F^{(d)}_{\rm mix}$\\
  \hline
    2 &    0.7116 &    1.0000 &    1.0000 &    1.0000 &  0.7116\\
    3 &    0.6580 &    1.0000 &    0.9956 &    0.9956 &  0.6580\\
    4 &    0.5788 &    0.9948 &    0.9947 &    0.9998 &  0.6369\\
    5 &    0.5616 &    0.9957 &    0.9950 &    0.9993 &  0.6183\\
   10 &    0.5341 &    0.9977 &    0.9969 &    0.9993 &  0.5769\\
   11 &    0.5317 &    0.9978 &    0.9972 &    0.9993 &  0.5726\\
   20 &    0.5206 &    0.9988 &    0.9984 &    0.9996 &  0.5513\\
   21 &    0.5199 &    0.9988 &    0.9984 &    0.9996 &  0.5499\\
  100 &    0.5076 &    0.9997 &    0.9997 &    0.9999 &  0.5212\\
  101 &    0.5075 &    0.9997 &    0.9997 &    0.9999 &  0.5211\\
  \hline
\end{tabular}
\end{table}

By analyzing the Wigner functions in Figs.~2 and~4 for increasing
$\alpha$, one can interpret the state $\ket{\alpha}_d$ at the
midpoint of the quasiperiod $T_d$ as a result of the interference
of a single QCS $\ket{\alpha}_d$ with its reflection
$|-\alpha\rangle_d$ from the Fock state $\ket{d}$ at the boundary
of the Hilbert space.

\subsection{Optical tomograms for cat states}

The optical tomogram $w_{\psi}(q,\theta)$ is the marginal
distribution of the Wigner function $W_{\psi}(q,p)$ for a given
state $\ket{\psi}$ of the quadrature component $q$ rotated by
angle $\theta$ in the quadrature phase space~\cite{SchleichBook}:
\begin{equation}
w_{\psi}(q,\theta) =\int_{-\infty}^{\infty} W_{\psi}\left(q\cos\theta
-p\sin\theta,q\sin\theta +p\cos\theta\right) dp. \label{tomogram1}
\end{equation}
Tomograms are directly measurable in homodyne detection, which
enable an indirect reconstruction of the Wigner function. So, one
can match experiment with theory. This particular feature of
tomograms makes them useful. Recently, Filippov and
Man'ko~\cite{Filippov11} obtained a closed form analytic
expression for the optical tomogram of any qudit superposition
state, given by Eq.~(\ref{eq:state1}), as
\begin{eqnarray}
\label{tomogram} w_{\psi}(q,\theta)  =
\frac{e^{-q^{2}}}{\sqrt{\pi}}\sum_{n=0}^{d-1}
\left[\frac{|c_{n}|^{2}}{2^{n}n!}H_{n}^{2}(q)\right.
+\frac{|c_{n}|}{\sqrt{2^{n}n!}}H_{n}(q)
\nonumber\\
  \times \sum_{k=n+1}^{d-1}
  \left.\frac{|c_{k}|\cos\left[\left(n-k\right)\theta-\phi_{n}+\phi_{k}\right]}
  {\sqrt{2^{k-2}k!}}H_{k}(q)\right],
\end{eqnarray}
where $c_{j}=|c_{j}|e^{i\phi_{j}}$ and $H_{j}(q)$ is the Hermite
polynomial of degree $j$ ($j=n,k$). We have used
Eq.~(\ref{tomogram}) to obtain tomograms of the QCS \NCS and \LCS.

Figures~8(b,c) and 9 show a few examples of the tomograms for the
low-dimensional Schr\"odinger cat states
$|\alpha=T_d/2\rangle_{d}$ in comparison to
$|\beta=T_d/2\rangle_{d}$. In addition, Fig.~8(a) shows a
single-photon state $|\alpha=T_2/2\rangle_{2}=\ket{1}$, which can
be considered a singular ``cat'' state. It is seen for \LCS that
the tomograms have two main peaks (if the divided peaks at the
boundaries for $\theta=0$ and $2\pi$ are combined together) and
$2(d-2)$ smaller peaks, so altogether $2(d-1)$ peaks. The total
number of peaks of the tomograms for \NCS, in comparison to \LCS,
is more difficult to be estimated for arbitrary $d$ because, e.g.,
some peaks are not well separated [e.g., compare the smallest
peaks in Figs.~9(a,b)]. For \NCS, there are altogether four
outermost peaks on the left and right-hand sides independent of
the dimension $d$, and a few squeezed peaks between them depending
on $d$, as clearly seen in Fig.~9. Note that \NCS cannot be
precisely obtained by simply superimposing the tomograms for
$|\beta\rangle_{d}$ and $|-\beta\rangle_{d}$ (which is a
$\pi$-rotated tomogram for $|\beta\rangle_{d}$). This would
correspond to a tomogram for $\rho^{(d)}_{\rm mix}$, given by
Eq.~(\ref{rho_mix}), for which the corresponding fidelity
$F^{(d)}_{\rm mix}$, given by Eq.~(\ref{Fmix}), is quite low, as
shown in Table~I.

The tomograms for $\ket{\psi}= |\alpha=T_d/2\rangle_{d}$ are
\emph{approximately} symmetric with respect to reflection along
the axes $q=0$ and $\theta=\pi$ (in addition to the symmetry along
$\theta=0$), i.e., $w_{\psi}(q,\theta)\approx w_{\psi}(-q,\theta)$
and $w_{\psi}(q,\pi+\theta)\approx w_{\psi}(q,\pi-\theta)$. Only
the latter symmetry is observed for
$\ket{\psi}=|\beta=T_d/2\rangle_{d}$ as seen in Figs.~8(d,e,f).
Note that the imperfections of the symmetries come from the
imperfect cat-state generation (i.e., $|\alpha\rangle_{d}$ is not
exactly equal to $ |\alpha_{\pm}\rangle_{d}$ for $\alpha=T_d/2$
with $d>3$) and, more importantly, from the interference in phase
space, which means that the tomograms (and the corresponding
Wigner functions) of any superpositions of states
$|\alpha\rangle_{d}$ and $|-\alpha\rangle_{d}$ are more asymmetric
than their mixtures.

\begin{figure}
\fig{ \includegraphics[scale=0.45]{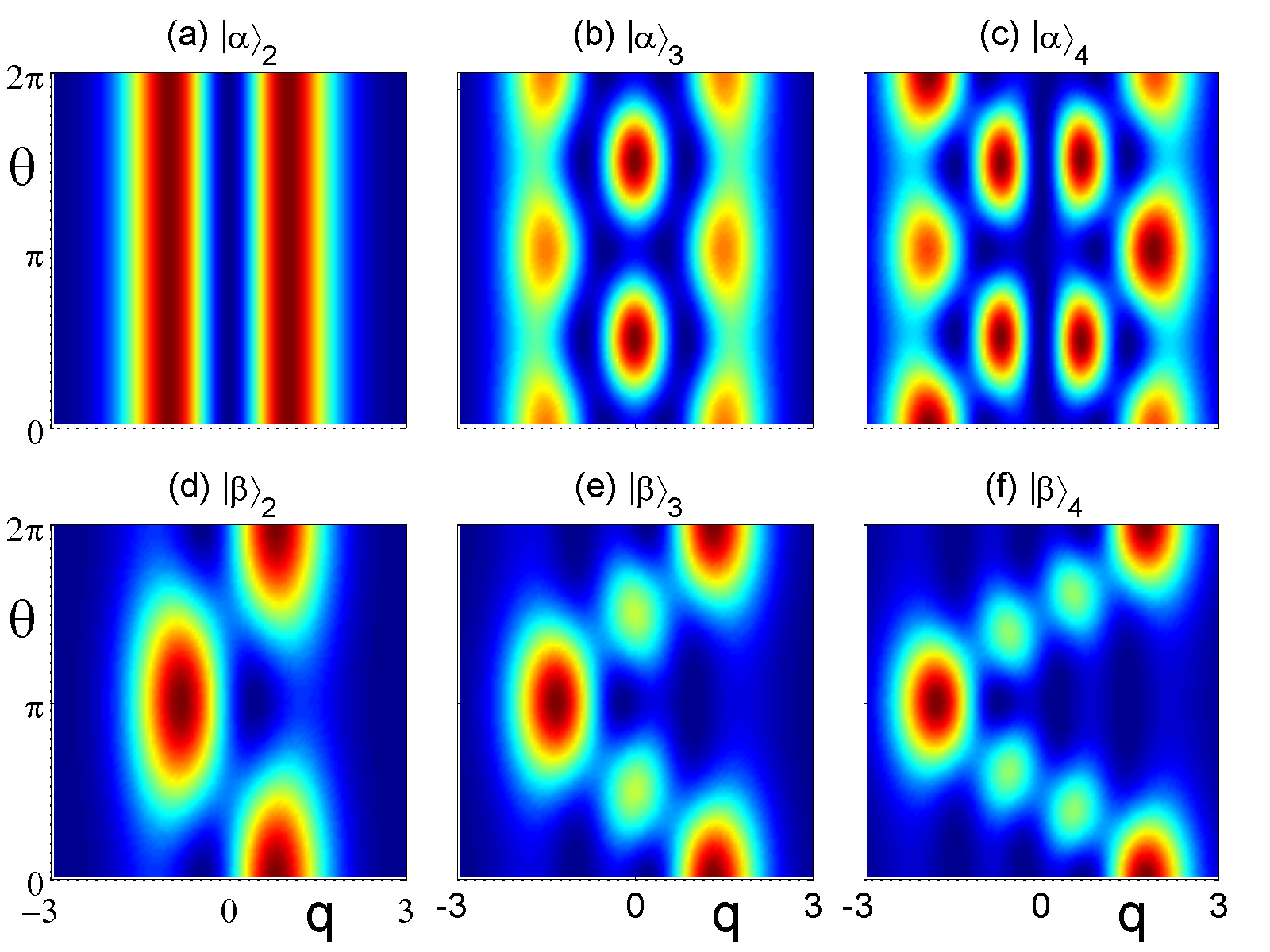}}

\caption{(Color online) Optical tomograms for the QCS \NCS (a,b,c)
and \LCS (d,e,f) for $\alpha=\beta=T_d/2$ with $d=2,3,4$. Dark
blue (dark orange) regions show zero (maximum) values. The upper
row tomograms correspond to (a) the single-photon Fock state, (b)
the even QCS (the Schr\"odinger male cat state)
$\ket{\alpha}_3\approx \ket{\alpha_+}_3\approx \ket{\beta_+}_3$,
and (c) odd QCS (female cat state) $\ket{\alpha}_4\approx
\ket{\alpha_-}_4\approx \ket{\beta_-}_4$. The tomograms are
$2\pi$-periodic in $\theta$, thus the divided peaks near
$\theta=0,2\pi$ should be understood as combined together. }
\end{figure}
\begin{figure}
\fig{ \includegraphics[scale=0.45]{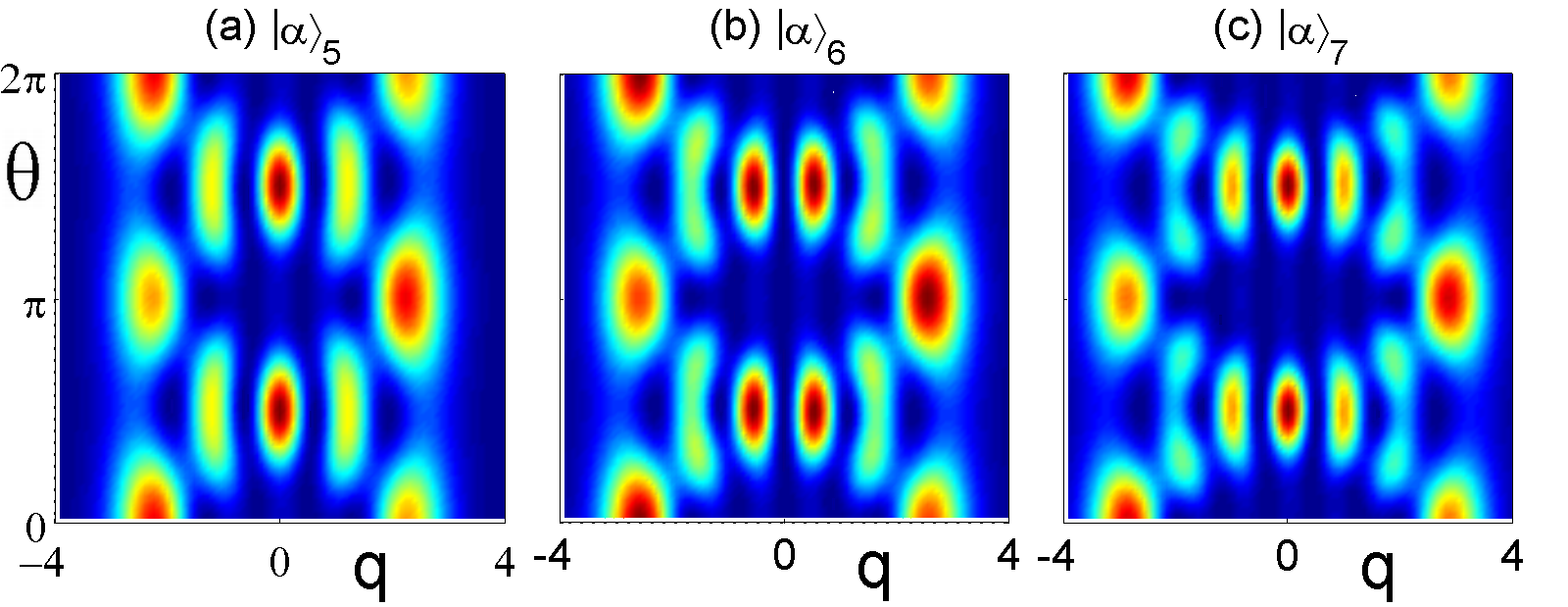}}

\caption{(Color online) Optical tomograms for the even (for
$d=5,7$) and odd ($d=6$) QCS $\ket{\alpha=T_d/2}_d$. }
\end{figure}

\subsection{Nonclassicality of the cat states}

A quantum state can be considered {\em nonclassical} if its
Glauber-Sudarshan $P$ function cannot be interpreted as a {\em
classical} probability density, i.e., it is
nonpositive~\cite{VogelBook}. In particular, if the $P$ function
is more singular than the Dirac delta function then it is also
nonpositive~\cite{Miran10}. Thus, any qudit state (including our
QCS), which is not the vacuum state, is nonclassical as any finite
superposition of Fock states is nonclassical.

There are various measures and criteria (witnesses) of
nonclassicality of optical states~\cite{VogelBook,Miran10}.
Formally the best measures are those based directly or indirectly
on the $P$ function. However, due to the singularity of the
$P$~function, they are not operationally useful except for some
very special states. Thus, we use an operational parameter (or a
quantitative witness) of nonclassicality based on the Wigner
function.

Here, in the analysis of the qudit Schr\"odinger cat states, we
apply the nonclassical volume, which is a quantitative parameter
of the amount of nonclassicality of a given quantum state based on
the Wigner function~\cite{Kenfack04}. In this particular measure,
the volume of the negative part of the Wigner function is
considered as an indicator (or parameter) of nonclassicality. To
be precise, the nonclassical volume is defined as a doubled volume
of the integrated negative part of the Wigner function of a
quantum state $|\psi\rangle$~\cite{Kenfack04}:
\begin{equation}
\delta(\ket{\psi})=\int_{-\infty}^{\infty}\!\int_{-\infty}^{\infty}\left|W_{\psi}\left(q,p\right)\right|dq
dp-1,
\end{equation}
where $W_{\psi}\left(q,p\right)$ is the Wigner function of a
quantum state $|\psi\rangle.$ A nonzero value of
$\delta(|\psi\rangle)$ implies that the given state $|\psi\rangle$
is nonclassical. For example, the vacuum is a classical state so
$\delta(\ket{0})=0$, while the single-photon Fock state has the
nonclassical volume equal to $\delta(\ket{1})=4 e^{-1/2}-2\approx
0.426$~\cite{Kenfack04}.

By analyzing Fig.~10, which shows the nonclassical volume
$\delta$, one can conclude that, at least for small $d$, the
following properties hold:
 (a) $\delta\left(\ket{\alpha=\tfrac12 T_{d+1}}_{d+1}\right)> \delta\left(\ket{\alpha=\tfrac12T_{d}}_{d}\right),$
 (b) $\delta\left(\ket{\beta}_{d+1}\right)> \delta\left(\ket{\beta}_{d}\right)$ if $|\beta|\gg0,$
 while (c) $\delta\left(\ket{\beta}_{d+1}\right)< \delta\left(\ket{\beta}_{d}\right)$
  if $|\beta|\approx0 $,
 and (d) analogously $\delta\left(\ket{\alpha}_{d+1}\right)<
 \delta\left(\ket{\alpha}_{d}\right)$
if $|\alpha|\approx0$. Moreover,
 (e) $\delta(\ket{\alpha}_{d})> \delta(\ket{\beta}_{d})$ if $|\alpha|=|\beta|\le \tfrac12 T_d.$

It is seen in Fig.~10(a) that $\delta(\ket{\alpha}_{2})$ reaches
its maximum value of $4 e^{-1/2}-2$ for $\alpha=T_{2}(n+1/2)$ with
$n=0,1,...$, which corresponds to the generation of the Fock state
$\ket{1}$ [see also Fig.~1(c)]. For higher $d$, the local maxima
of $\delta(\ket{\alpha}_{d})$ are also reached for
$\alpha=T_{d}(n+1/2)$, which corresponds to the generation of the
even and odd cat states. So, in terms of the nonclassical volume,
the most nonclassical QCS \NCS, for a given $d$, are the cat
states. This fact also justifies our choice of $\alpha=T_d/2$ for
the construction of the tomograms shown in Figs.~8 and~9.

It is seen in Fig.~10(c) that the range of $\alpha=\beta$ for
which $\delta(\ket{\alpha}_{d})\approx
\delta(\ket{\beta}_{d})\approx 0$ increases with $d$, also as a
fraction of $T_d$, for both types of QCS. This indirectly shows
that these QCS tend to the conventional Glauber coherent states
for $|\alpha|=|\beta|\ll d$.

These and other quantifiers and witnesses were also analyzed in
the context of the generation of standard infinite-dimensional
Schr\"odinger cat states and their quantum-to-classical transition
by, e.g., Paavola \etal~\cite{Paavola11}. Their analysis of
nonclassicality includes: (1) a nonclassical depth based on the
$s$-parametrized generalization of the Glauber-Sudarshan and
Wigner functions, (2) the highest point of the interference
fringes of the Wigner function, (3) the Vogel nonclassicality
criterion based on the matrices of moments of annihilation and
creation operators, and (4) the Klyshko criterion based on the
photon-number distribution in addition to (5) the nonclassical
volume studied here. Numerous other nonclassicality parameters,
which can also be applied in this context, are listed in, e.g.,
Ref.~\cite{Miran10}.

\begin{figure}
\fig{
\includegraphics[scale=0.31]{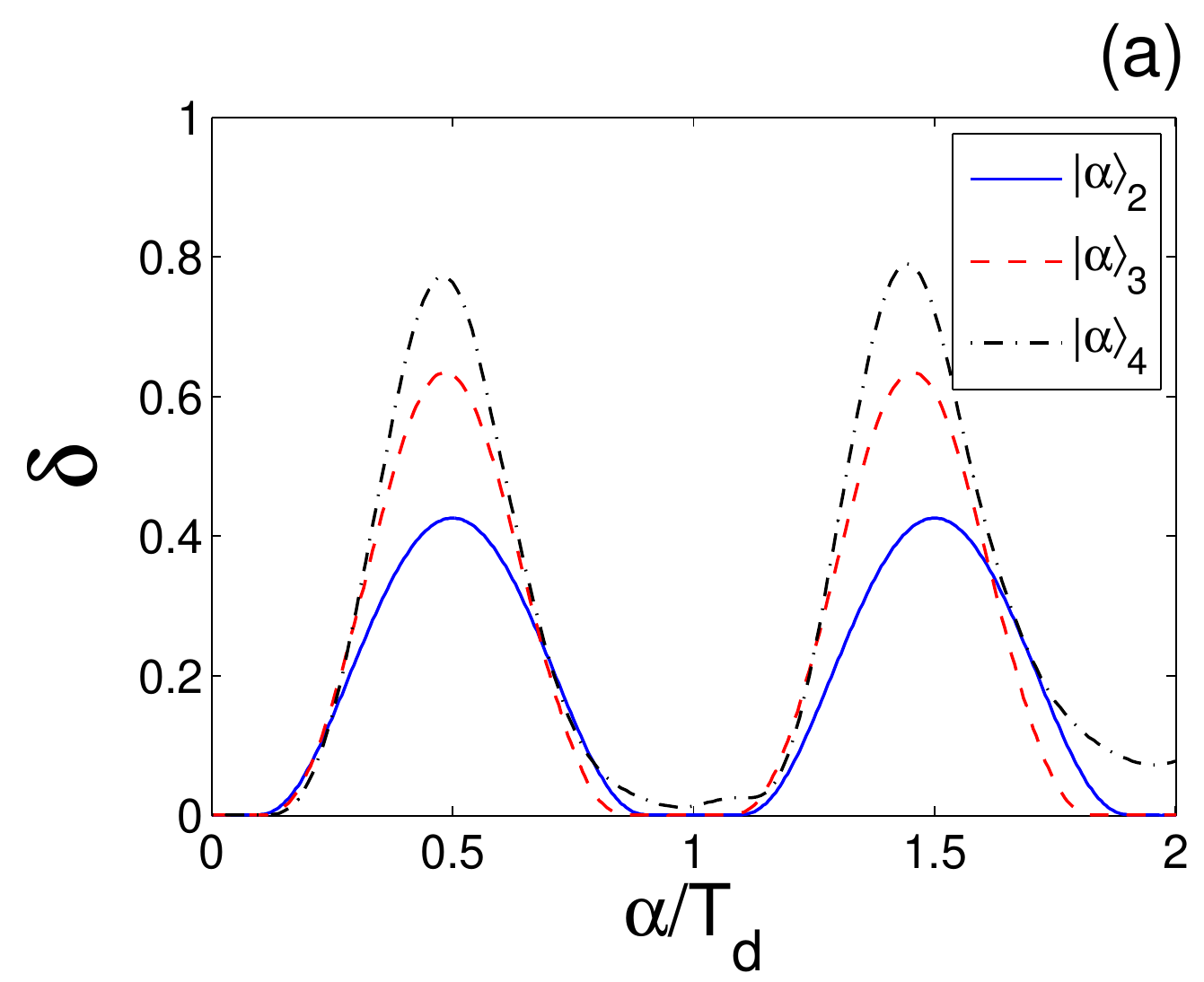}
 \includegraphics[scale=0.31]{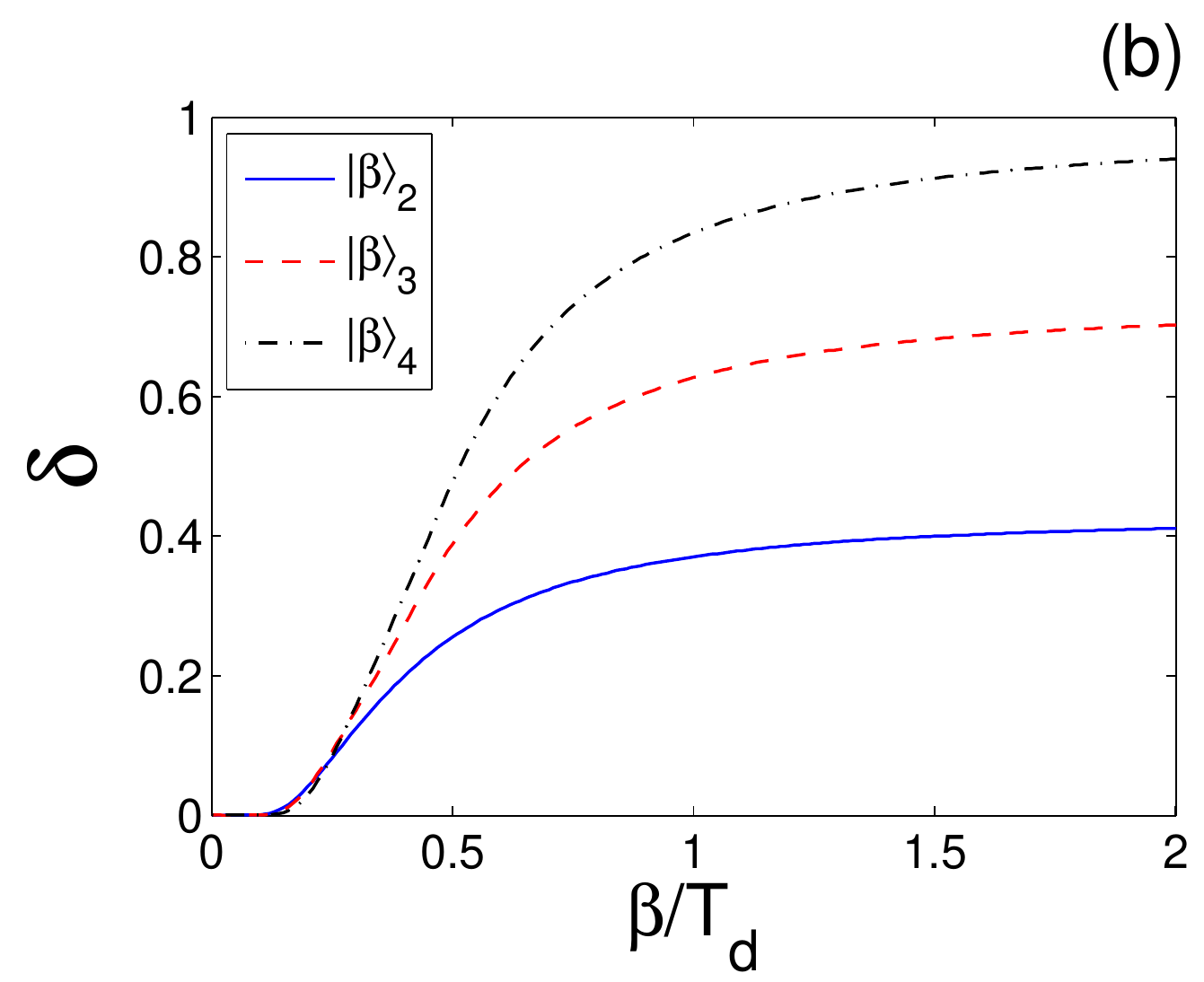}}

\fig{
  \includegraphics[scale=0.31]{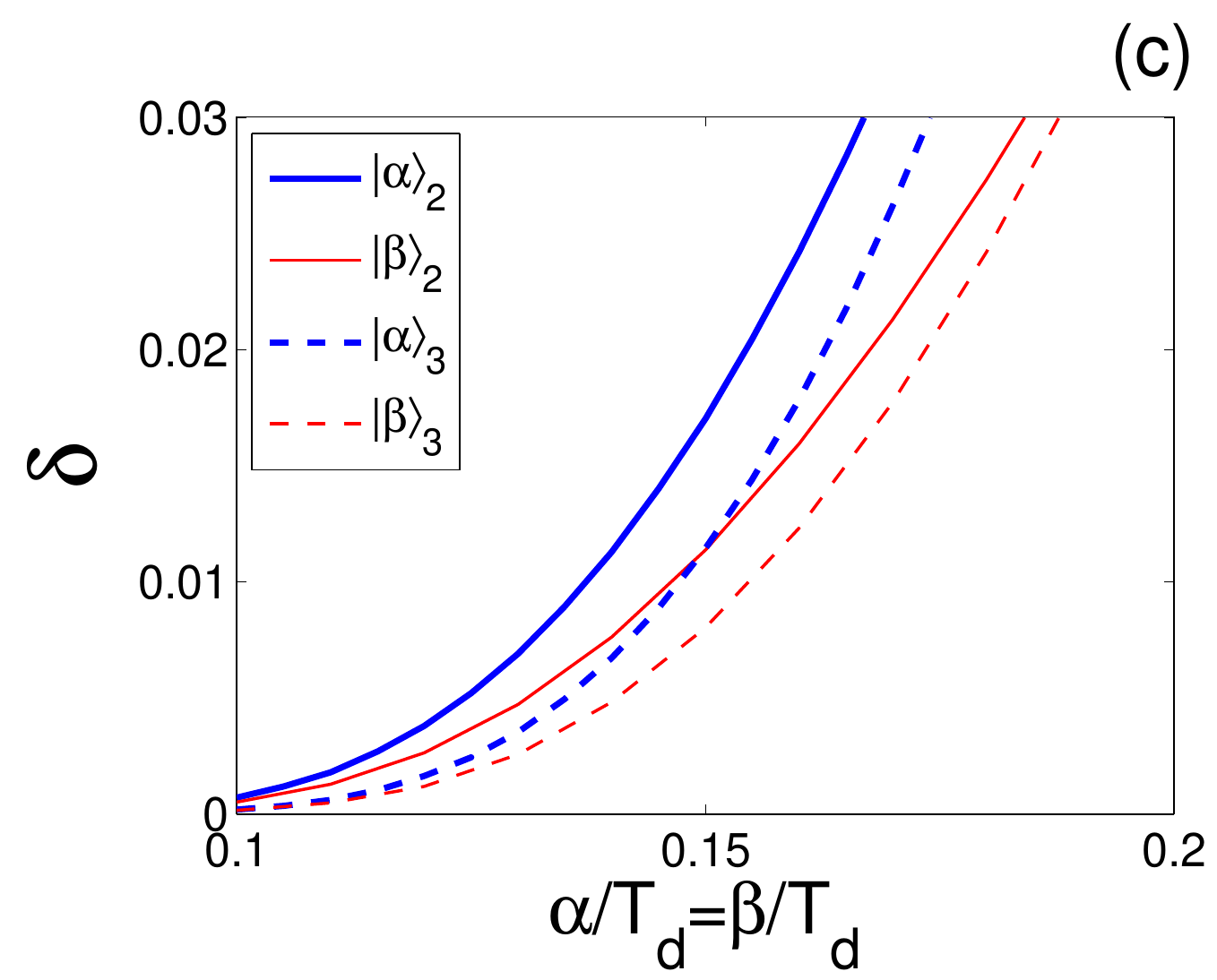}
   \includegraphics[scale=0.31]{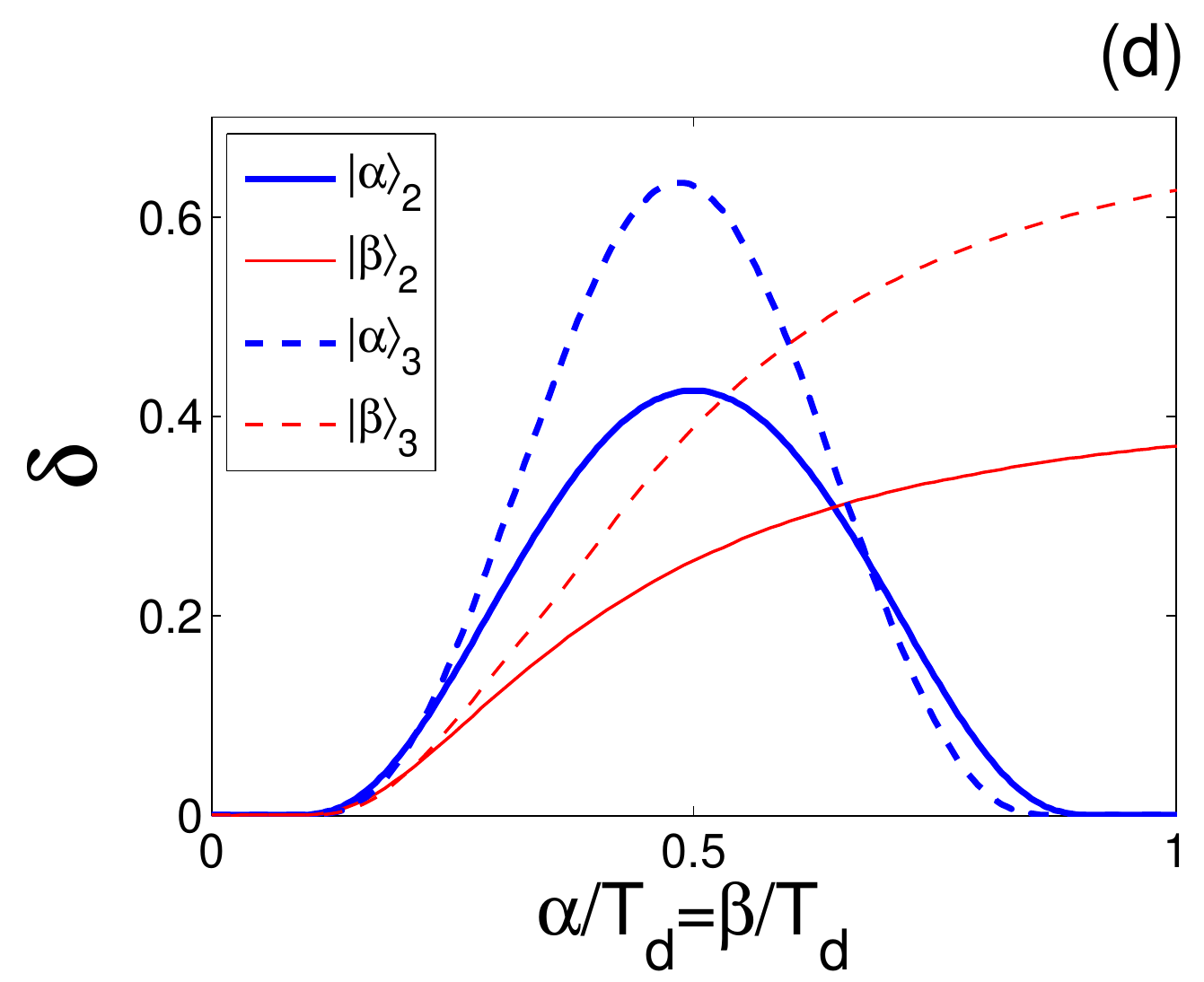}}

\caption{(Color online) Variation of the nonclassical volume
$\delta(\ket{\psi})$ with the real amplitudes $\alpha$ and $\beta$
as a fraction of the quasiperiods $T_d$ for the QCS
$\ket{\alpha}_{d}$ and $|\beta\rangle_{d}$ with $d=2,3,4$. }
\end{figure}

\section{Conclusions}

We compared properties of two kinds of qudit (or $d$-level) CS:
(i) \NCS defined by the action of the qudit displacement operator
on the vacuum and (ii) \LCS defined by the Poissonian expansion in
Fock states truncated at the state $\ket{d-1}$. In the
infinite-dimensional limit of the Hilbert space or, practically,
if $|\alpha|=|\beta| \ll d$, these two QCS go into the same
conventional Glauber CS. Also the states are equivalent for the
qubit case (i.e., for $d=2$). However, for other cases, the QCS
\NCS and \LCS exhibit distinctly different properties. The crucial
difference between these two types of QCS is that the state \NCS
with increasing $\alpha=\beta$ exhibits periodic (for $d=2,3$) or
quasiperiodic (for $d>3$) reflections from the boundary states
$\ket{0}$ and $\ket{d-1}$ of the Hilbert space ${\cal H}^{(d)}$,
which we described as multiple bounce or a ping-pong effect. By
contrast, the QCS \LCS is not reflected from the boundaries of the
Hilbert space as $\beta$ increases, which produces no reflections
and no bouncing of the Wigner function. Although the
quasiperiodicity of the QCS \NCS was already discussed in
Refs.~\cite{Opatrny96,Leonski97b}, our phase-space description in
terms of the standard Wigner function shows these effects
especially clearly in terms of quantum interference in phase
space.

We have shown analytically that the QCS \NCS for $\alpha=T_d/2$
form macroscopically distinguishable superpositions of two qudit
CS. Thus, these Schr\"odinger cat states can be simply generated
by a direct displacement of the vacuum state in a qudit system.
The cat states can contain Fock states with only odd or even
photon numbers, depending on whether the qudit dimension $d$ is
even or odd, and thus referred to as the odd or even QCS,
respectively. We have interpreted this phenomenon as an
interference of a single CS \NCS with its reflection
$\ket{-\alpha}_d$ from the highest-energy Fock state $\ket{d-1}$
of the qudit Hilbert space.

Various experimental methods (see, e.g. Refs.~\cite{OpticalCats}
and references therein) have been developed for the generation of
quantum superpositions of two and more well-separated
quasiclassical states of light, referred to as the Schr\"odinger
cat and kitten states, respectively. In particular, it is well
known theoretically, and recently confirmed
experimentally~\cite{Kirchmair13}, that an initial CS in a Kerr
medium with the third-order nonlinear susceptibility can evolve
into a superposition of two~\cite{Yurke86,Milburn86,Tombesi87} or
more~\cite{Miran90} macroscopically distinct superpositions of
infinite-dimensional CS. Also the evolution of an initial coherent
state through a Kerr medium, described by a higher-order nonlinear
susceptibility, results in the production of Schr\"odinger
cat~\cite{Yurke86,Tombesi87} and kitten~\cite{Paprzycka92} states.
We note that Schr\"odinger cat states can also be produced in a
microwave cavity field via its coupling to a superconducting qubit
in circuit-QED systems~\cite{Liu05}, which under special
conditions can be modeled as a Kerr-type effect.

The generation of finite-dimensional even and odd cat states
discussed in this paper corresponds to a completely different
effect as based on simple displacements of the vacuum. The Kerr
effect, as shown in Fig.~1(a), was used only as an example of the
optical method for the Hilbert-space truncation.

It is worth noting that the QCS, for any nonzero $\alpha$ and
$\beta$, are not classical, in contrast to their
infinite-dimensional counterpart. It is known that any qudit state
different from the vacuum is nonclassical because any finite
superposition of Fock states is nonclassical, i.e., described by a
non-positive semidefinite Glauber-Sudarshan $P$~function. However,
so far no effort has been made to compare the nonclassical
properties of these two types of QCS. Keeping this in mind, here
we investigated the differences between the nonclassical
properties of the two types of QCS.

We have illustrated the nonclassical properties of the two types
of QCS by studying their photon-number statistics and the
nonclassical volume of the Wigner function, which is the
Kenfack-\.{Z}yczkowski quantitative parameter of
nonclassicality~\cite{Kenfack04}.

By showing similarities and clear differences of
finite-dimensional (nonclassical) and infinite-dimensional
(classical) systems depending on the complex parameters (such as
$\alpha$ and $\beta$) in comparison to the system dimension, one
can address fundamental questions of the quantum-to-classical
transition.

For the completeness of our phase-space description, we have also
presented optical tomograms of the QCS. These tomograms, which are
directly measurable in homodyne detection, enable the complete
reconstruction of the Wigner function.

We stress that the discussed QCS are not only of fundamental
theoretical interest, as they can be generated in optical systems
referred to as the linear and nonlinear quantum scissors (see
Secs.~II and~III).

We would like to emphasize that we studied the generation of
Schr\"odinger cat states in a \emph{general} finite-dimensional
bosonic system in which the displacement operation can be applied
to the ground state of the system. Figure~1 shows just a few
examples of optical realizations of such systems often studied in
the literature. Although these systems are theoretically appealing
because of their formal simplicity, we do not claim that they are
the easiest to be realized experimentally. Especially, when one
uses Kerr nonlinearities modeled by a $d$-photon anharmonic
oscillator, which is required in the system shown in Fig.~1(a) for
$d>2$. We are not aware of any direct experimental realization of
a pure $d$-photon anharmonic oscillator for $d>2$, although this
model was used in a number of theoretical works including the
classic articles by Yurke and Stoler~\cite{Yurke86}, and Tombesi
and Mecozzi~\cite{Tombesi87} on the Schr\"odinger cat generation.
By contrast, the system shown in Fig.~1(a) in the special case of
Kerr nonlinearity described by the two-photon  anharmonic
oscillator enables single-photon
blockade~\cite{Imamoglu97,Leonski94}, corresponding to the
generation of lossy two-dimensional CS $\ket{\alpha}_{2}$. This
effect has already attracted much attention and was demonstrated
in a number of experiments in cavity- and circuit-QED
setups~\cite{Birnbaum05, Faraon08, Hoffman11,Lang11}. Also the
systems shown in Figs.~1(b) and 1(c) were realized experimentally
as reported in, e.g., Refs.~\cite{Babichev03} (according to the
experimental proposal of Ref.~\cite{Ozdemir01}) and \cite{Reck94},
respectively.

We hope that this work can stimulate further interest in finding
applications of the QCS in quantum information-processing
(including quantum teleportation) with qudits and quantum
engineering with multiphoton blockades.

\begin{acknowledgments}
A.M. was supported by the Polish Ministry of Science and Higher
Education under Grant No. DEC-2011/03/B/ST2/01903. A.P. thanks the
Department of Science and Technology (DST), India, for support
provided through the DST project No. SR/S2/LOP-0012/2010 and he
also thanks the Operational Program Education for Competitiveness
-- European Social Fund project CZ.1.07/2.3.00/20.0017 of the
Ministry of Education, Youth and Sports of the Czech Republic.
F.N. is partially supported by the RIKEN iTHES Project, MURI
Center for Dynamic Magneto-Optics, JSPS-RFBR Contract No.
12-02-92100, Grant-in-Aid for Scientific Research (S), MEXT
Kakenhi on Quantum Cybernetics, and the JSPS via its FIRST
program.
\end{acknowledgments}

\appendix

\section{Simple examples of QCS}

Here, for clarity, we give two simple examples of the QCS \NCS,
showing their relation to the cat state generation.

Equation~(\ref{NQCS}) for $d=3$ simplifies to the qutrit
CS~\cite{Miran94}:
\begin{eqnarray}
\ket{\alpha}_3 & = & \frac{1}{3}[2+\cos(\sqrt{3}|\alpha|)]|0\rangle+\frac{1}{\sqrt{3}}e^{{\rm i}\phi_{0}}\sin(\sqrt{3}|\alpha|)|1\rangle\nonumber \\
 &  & +\frac{\sqrt{2}}{3}e^{2i \phi_{0}}[1-\cos(\sqrt{3}|\alpha|)]|2\rangle.
 \label{d3}
\end{eqnarray}
It is seen that the single-photon term exactly vanishes for
$\alpha=T_3/2=\pi/\sqrt{3}$, thus this superposition state reduces
exactly to the qutrit even CS, given by Eq.~(\ref{cat3}). The
Wigner functions for $\ket{\alpha}_3$ with various $\alpha$,
including $\alpha=T_3/2$, are shown in Fig.~4.

For $\alpha=\beta=\gamma=T_3/2$, one can calculate explicitly that
$\ket{\alpha}_3\approx[0.33,0,0.94]$, $\ket{\pm\beta}_3
\approx[0.32, \pm 0.58, 0.75]$, $\ket{\gamma}_3\approx[0.22,
-0.58, 0.78]$, and $\ket{\beta_+}_3\approx[0.37,0,0.93]$. Thus, it
is seen that $\ket{\alpha}_3\approx
\ket{\alpha_+}_3\approx\ket{\beta_+}_3$, which results in the
corresponding fidelities $\approx 1$ (see Table~I). This
conclusion can be drawn intuitively (but inaccurately) by
comparing the Wigner function $W(q,p;\ket{\alpha}_3),$ shown in
Fig.~4(c), with $W(q,p;\ket{\beta}_3)$, shown in Fig.~5(b),
roughly superimposed with  $W(q,p;-\ket{\beta}_3)$, which is
$W(q,p;\ket{\beta}_3)$ but rotated by $\pi$ in phase space
according to Eq.~(\ref{WignerSymmetry}).  As already mentioned,
such superimposing of plots corresponds to the Wigner function of
a mixed state $\rho^{(3)}_{\rm mix}$, given by
Eq.~(\ref{rho_mix}), clearly different from $\ket{\alpha}_3$ as
indicated by the relatively low fidelity $F^{(3)}_{\rm mix}=0.66$
(see Table~I).

For $d=4$, from Eq.~(\ref{NQCS}) one obtains the following
\emph{quartit} CS:
\begin{eqnarray}
\ket{\alpha}_{4}  =\frac{1}{2}  \sum_{k=1,2}
\Big\{\frac{1}{x^2_{k}}\cos y_{k}|0\rangle
 +\frac{e^{i \phi_{0}}}{x_{k}}
\sin y_k|1\rangle\hspace{8mm}\nonumber \\
 + (-1)^{k} \frac{e^{2i \phi_{0}}} {\sqrt{3}}\cos y_k|2\rangle
 + (-1)^{k}\frac{e^{3i \phi_{0}}}{x_{k}}\sin y_k|3\rangle\Big\} ,
 \label{d4}
\end{eqnarray}
where $x_{1,2}=x_{1,2}^{(4)}=\sqrt{3\pm\sqrt{6}}$ are the roots of
${\rm He}_4(x)$ and $y_k=x_{k}|\alpha|$. To show that this state
for $\alpha=T_4/2$ is close to the quartit odd CS, it is enough to
calculate the contributions of the Fock states $|0\rangle$ and
$|2\rangle$, which are $|\langle 0\ket{\alpha}_4|^{2}=0.0004$ and
$|\langle 2\ket{\alpha}_4|^{2}=0.0048$. These contributions are
clearly negligible, as also shown in Fig.~7(b). The Wigner
function and its tomograms for this cat state are shown in
Figs.~6(a) and 8(c) in comparison to the cat states generated in
the Hilbert spaces of other dimensions.

An explicit calculation for $\alpha=\beta=\gamma=T_4/2$ leads to
$\ket{\alpha}_4\approx[0.02,0.47,-0.07,0.88]$ [see Fig.~6(a)],
$\ket{\pm\beta}_4\approx[0.18,\pm0.38,0.57,\pm0.70]$ [see
Fig.~6(d)], $\ket{\gamma}_4\approx[-0.15,0.33,-0.68,0.64]$, and
$\ket{\beta_-}_4\approx[0,0.48,0,0.88]$. Thus, we see that
$\ket{\alpha}_4\approx \ket{\alpha_-}_4\approx \ket{\beta_-}_4$
resulting in a fidelity close to 1 (as shown in Table~I). This
conclusion can be drawn also for other dimensions $d$ as seen by
comparing Figs.~6(a,b,c) with Figs.~6(d,e,f), respectively.


\end{document}